\PassOptionsToPackage{unicode}{hyperref}
\PassOptionsToPackage{hyphens}{url}
\PassOptionsToPackage{dvipsnames,svgnames,x11names}{xcolor}
\documentclass[
  12pt]{article}

\usepackage{amssymb}
\usepackage{mathrsfs}
\usepackage{iftex}
\usepackage{multirow}
\usepackage{dsfont} 
\usepackage{graphicx,psfrag,epsf}
\usepackage{enumerate}
\usepackage{array}
\usepackage{url} 
\usepackage{amsmath}
\usepackage{tcolorbox}
\usepackage{subcaption}
\usepackage{lscape}
\usepackage{bm}
\usepackage{hyperref}
\usepackage{arydshln}
\usepackage{algorithm}
\usepackage{algpseudocode}
\usepackage{enumitem}
\usepackage{caption}
\usepackage{xr-hyper}

\let\hat\widehat

\newtheorem{remark}{Remark}[section]
\newcommand{\cN}{\mathcal{N}}
\newcommand{\cL}{\mathcal{L}}
\newcommand{\RR}{\mathbb{R}}

\newcommand{\argmax}{\mathop{\mathrm{argmax}}}

\def\T{{ \intercal }}

\makeatletter
\def\fps@figure{htbp}
\makeatother

\makeatletter

\usepackage{xcolor}
\usepackage[]{natbib}
\usepackage{bookmark}

\IfFileExists{xurl.sty}{\usepackage{xurl}}{} 
\hypersetup{
  pdftitle={Title},
  pdfauthor={Author 1; Author 2},
  pdfkeywords={3 to 6 keywords, that do not appear in the title},
  colorlinks=true,
  linkcolor={blue},
  filecolor={Maroon},
  citecolor={Blue},
  urlcolor={Blue},
  pdfcreator={LaTeX via pandoc}}

\begin{document}

\def\spacingset#1{\renewcommand{\baselinestretch}%
{#1}\small\normalsize} \spacingset{1}



  \title{\bf  Multidimensional Item Response Theory under General Latent Distributions}
  \author{Chengyu Cui$^{1*}$, Taoyi Chen$^1$\footnote{Equal contribution}, Chun Wang$^2$, and
Gongjun Xu$^1$, \vspace{.3cm}\\
{\small $^1$ \it Department of Statistics, University of Michigan}\vspace{.3cm}\\
    {\small $^2$ \it College of Education, University of Washington}
    }
    \date{}
  \maketitle

\medskip
\spacingset{1.6}
\begin{abstract}
Multidimensional item response theory (MIRT) provides an important psychometric framework for modeling how multiple latent traits jointly influence observed item responses. In most existing estimation procedures, the latent trait distribution is assumed to be Gaussian. Although computationally convenient, this assumption can be restrictive in many applications where the latent distribution exhibits skewness, heavy tails, or multimodality. More importantly, misspecifying the latent distribution may bias the estimation of item parameters and latent traits. To address this limitation, we propose a data-driven flow-based framework for MIRT models that can capture a broad class of non-Gaussian latent distributions. The proposed approach represents the latent distribution as an invertible transformation of a simple base distribution. For efficient estimation, we further introduce a conditional flow as a function of both the observed response and the noise to approximate the posterior distribution. Under this framework, the item parameters, latent distribution, and posterior approximation can be learned jointly. Comprehensive simulation studies show that the proposed method improves item-parameter and latent-trait recovery when the true latent distribution is non-normal. An application to a personality dataset further illustrates the practical utility of the proposed framework for modeling complex latent trait distributions in large-scale data.
\end{abstract}

\spacingset{1.8}


\captionsetup[figure]{width=1\linewidth}
\section{Introduction}

Multidimensional item response theory (MIRT) is a framework for modeling the relationship between observed item responses and latent traits used to represent unobserved proficiencies, abilities, or psychological factors. By allowing multiple latent traits to jointly influence response behavior, MIRT offers an interpretable and flexible tool for analyzing complex latent constructs in educational measurement, psychology, and the behavioral sciences. Owing to these features, this modeling framework has become an important research topic in psychometrics, particularly for studying large-scale data with complex latent structures~\citep{reckase2009}. 

Estimating MIRT models is both statistically and computationally challenging because the latent traits are unobserved. A widely used approach is marginal maximum likelihood estimation, in which the marginal likelihood is obtained by integrating the complete-data likelihood with respect to the latent trait distribution. The model parameters are then estimated by maximizing this marginal likelihood. A major challenge in this approach is that the required integrals become increasingly difficult to evaluate as the dimension of the latent trait vector increases. Classical approaches approximate these integrals using numerical methods such as Gauss--Hermite quadrature \citep{Bock_Aitkin_1981} or Laplace approximation \citep{Lindstrom1988NewtonRaphsonAE}. However, quadrature-based methods scale poorly when latent dimensions increase, while Laplace approximations may lose accuracy when the posterior distribution is highly skewed or otherwise poorly approximated by a local Gaussian form.
To improve scalability, a variety of simulation-based and stochastic estimation methods have been developed. These include Monte Carlo expectation-maximization algorithms \citep{mcculloch1997maximum}, stochastic EM procedures \citep{von2010stochastic}, and the Metropolis--Hastings Robbins--Monro (MHRM) algorithm \citep{cai2010high}. More recently, \citet{zhang2020improved} proposed a stochastic EM algorithm for item factor analysis using a Gibbs sampler in the stochastic E-step. Moreover, variational methods provide another important class of scalable approximations by replacing the intractable posterior distribution of the latent traits with a tractable variational approximation and optimizing an evidence lower bound to the marginal likelihood \citep{rijmen2013fitting,cho2021gaussian, cho2024regularized, cui2024variational,XiaoWangXu2024}. These methods substantially reduce the computational burden and are particularly attractive for large-scale settings.

Despite these developments in improving computational efficiency, many existing methods still rely on a multivariate Gaussian assumption for the latent trait distribution, which may provide an overly simplified representation of the underlying latent structure. While this assumption is convenient for estimation, it can be restrictive in practice. In applications, latent trait distributions can deviate from normality because of heterogeneous respondent populations, ceiling or floor effects, subgroup structure, unobserved covariates, or selection mechanisms. Such departures may give rise to skewed, heavy-tailed, or multimodal latent distributions~\citep{woods2006item}. When the assumed latent distribution differs from the true distribution, item parameter estimates and latent trait estimates may be biased~\citep{li2018summed, wang2018robustness}, substantially limiting the practical relevance of MIRT modeling.

A growing body of work has considered alternative specifications for the latent trait distribution beyond the standard Gaussian assumption. These include tests and models for heteroscedastic latent traits~\citep{molenaar2012heteroscedastic}, parametric non-Gaussian specifications such as skew-normal latent distributions \citep{santos2013multiple}, semiparametric density estimation using Ramsay-curve representations~\citep{monroe2014estimation}, and Bayesian methods based on flexible nonnormal density approximations~\citep{zhang2021bayesian}. 
However, these approaches have several limitations. Parametric non-Gaussian models can capture specific departures from normality, such as skewness, but may still be restrictive for more complex latent distributions, including those with local asymmetry or multimodality. For semiparametric and mixture-based approaches, performance often depends sensitively on choices of basis functions, smoothing parameters, mixture components, or other tuning specifications. Moreover, Bayesian approaches typically require computationally intensive sampling procedures, making them less scalable for high-dimensional latent traits or large-scale response data. As a result, developing a framework that can flexibly approximate complex latent trait distributions while remaining computationally scalable remains an important problem in MIRT estimation.

In this paper, we propose a data-driven flow-based framework for estimating MIRT models that accommodate a broad class of latent trait distributions. Normalizing flows have become a popular tool in generative modeling for constructing complex distributions by transforming samples from a simple base distribution through a sequence of invertible maps~\citep{rezende2015variational, lipman2022flow}. Inspired by this idea, we model the latent trait distribution as an invertible transformation of a Gaussian random vector. This formulation allows tractable density evaluation through the change-of-variables formula while providing sufficient flexibility to capture increasingly complex latent distributions as the expressiveness of the transformation increases.

Unlike standard uses of normalizing flows as unconditional density models, our framework uses normalizing flows to model both the latent prior and the posterior approximation. Specifically, for posterior inference, we introduce a conditional normalizing flow that maps a Gaussian noise vector, together with the observed responses, into the latent trait space. This construction yields a flexible posterior family that can capture complex, data-dependent uncertainty in the latent traits. It also distinguishes our approach from standard variational autoencoders~\citep{kingma2014auto}, in which the encoder often parameterizes a relatively simple approximate posterior family, such as a Gaussian distribution with data-dependent parameters. We further develop a stochastic optimization algorithm that jointly estimates the item parameters, prior-flow parameters, and posterior-flow parameters. To demonstrate the effectiveness of the proposed framework, we conduct a series of numerical studies under non-Gaussian latent distribution designs. We also apply the proposed method to a Big Five personality dataset, illustrating its practical utility for modeling complex latent trait distributions in large-scale psychometric data.




The remainder of the paper is organized as follows. Section~\ref{sec_setup} introduces the model setup and the flow-based specification of the latent trait distribution. Section~\ref{sec_proposed_method} develops the proposed estimation framework, including the conditional flow posterior and the stochastic optimization algorithm. Section~\ref{sec_simu} evaluates the proposed method through simulation studies. Section~\ref{sec_data} presents an empirical application to the personality dataset. Section~\ref{sec_conclude} concludes with a discussion and directions for future research.

\section{Problem Setup}\label{sec_setup}
We consider the multidimensional two-parameter logistic model (M2PL), a widely used compensatory model that allows multiple latent traits to influence each response. Suppose there are $N$ individuals. For each individual $i\in[N]:=\{1,2\dots, N\}$, we observe a $J$-dimensional binary response vector $Y_i = (Y_{i1},\dots, Y_{iJ})^\T$. Let $\theta_i\in\RR^{K}$ denote the $K$-dimensional latent trait for individual $i$, representing multiple facets of the underlying trait. The M2PL model specifies the conditional distribution of each response $Y_{ij}$ given $\theta_i$ as 
\begin{equation*}
    \Pr(Y_{ij}=1\mid \theta_i) \; =\; \sigma(\alpha_j^\top \theta_i - b_j),
\end{equation*}where $\sigma(x) := \{1+\exp(-x)\}^{-1}$ is the logistic function, $\alpha_j\in\mathbb{R}^K$ is the item discrimination
vector, and $b_j\in\mathbb{R}$ is the item difficulty parameter. The discrimination parameter $\alpha_j$ characterizes how strongly the response to item $j$ is associated with each latent trait, while the difficulty parameter $b_j$ controls the overall tendency of individuals to answer item $j$ correctly. Define the collection of the item parameters as $\psi = \{a_j,b_j\}_{j\in[J]}$. Following the literature~\citep{reckase2009,embretson2025item}, we assume local independence, meaning that conditional on the latent trait $\theta_i$, the item responses $\{Y_{ij}\}_{j\in[J]}$ are mutually independent. Under this assumption, the conditional response probability factors over items, and the marginal log-likelihood for the item parameters $\psi$ is therefore given by
\begin{equation}\label{eq_marginallk_1}
    \cL(\psi\mid Y) \;=\; \sum_{i=1}^N\log p_{\psi}(Y_i),
\end{equation}
where letting $\pi(\theta_i)$ denotes the density of the latent trait distribution, the marginal response probability $p_{\psi}(Y_i)$ is
\begin{equation}\label{eq_marginallk_2}
\begin{aligned}
    p_{\psi}(Y_i) = &\,\int_{\theta_i\in\RR^{K}}p_{\psi}(Y_i,\theta_i)d\theta_i \\
    = &\,\int_{\theta_i\in\RR^K}\prod_{j=1}^{J} \sigma(\alpha_j^\top \theta_i - b_j)^{Y_{ij}} \left\{1-\sigma(\alpha_j^\top \theta_i - b_j)\right\}^{1-Y_{ij}}\pi(\theta_i)d\theta_i\;. 
\end{aligned}
\end{equation}Here, $p_{\psi}(Y_i,\theta_i)$ is the joint density or probability mass of $Y_i$ and $\theta_i$ under the model. In many marginal estimation methods, the latent trait vector is assumed to follow a multivariate Gaussian distribution~\citep{reckase2009,hartig2009multidimensional}, i.e., $\pi(\cdot)$ is specified as
\[
p_{\cN}(\theta\mid \mu,\Sigma) \,=\, (2 \pi)^{-K/2} |\Sigma|^{-1/2} \exp\big\{ -1/2 (\theta_i - \mu)^T \Sigma^{-1} (\theta_i - \mu) \big\},
\] where $\mu$ is the latent mean vector and $\Sigma$ is the latent covariance matrix.
However, empirical latent trait distributions may deviate from normality, and misspecifying the latent distribution can bias item calibration and trait recovery~\citep{li2018summed, wang2018robustness}. This motivates modeling strategies allowing non-normal latent distributions. 

In this work, we take a flexible and expressive approach inspired by recent flow-based generative modeling, which constructs complex distributions by transforming samples from simple base distributions through invertible maps~\citep[see, e.g.,][]{rezende2015variational,lipman2022flow}. Specifically, we assume that the latent trait vector for individual $i$ is generated as
\begin{equation}
    \theta_i \,=\, h_{\eta}(\varepsilon);\quad \varepsilon_i\sim \cN(0,I_K),\label{eq_raw_prior_flow}
\end{equation}where $I_K$ is the $K\times K$ identity matrix and $h_\eta(\cdot)$ is an invertible transformation parameterized by flow parameters $\eta$. This formulation allows us to efficiently sample $\theta_i$ by first drawing $\varepsilon_i$ from the standard Gaussian base distribution and then mapping it through $ h_\eta(\cdot)$. The induced latent trait density can be evaluated by the change-of-variables formula:
\begin{equation}
    \pi_\eta(\theta) \;= \; p_{\cN}\big(h_\eta^{-1}(\theta)\mid 0,I_K\big)\; \big| \det \mathcal{J}_{h_\eta^{-1}}(\theta) \big|.\label{eq_change_of_var}
\end{equation}Here, $\mathcal{J}_{h_\eta^{-1}}(\theta)$ denotes the Jacobian matrix of the inverse transformation evaluated at $\varepsilon$. In our implementation, $h_\eta$ is modeled by a neural network. As the depth and width of the neural network increase, the induced density $\pi_\eta(\cdot)$ can capture increasingly complex latent trait distributions.

Given this flow-based latent distribution, we estimate the full parameter set, consisting of the item parameters $\psi$ and the flow parameters $\eta$, by marginal maximum likelihood:
\begin{equation}\argmax_{\psi,\eta}\;\sum_{i=1}^N\log\int_{\theta_i\in\RR^K}\Pr(Y_i\mid\theta_i,\psi) \pi_{\eta}(\theta_i)d\theta_i\, .\label{eq_mmle}
\end{equation}
This estimation problem is challenging for two reasons. First, the maximization problem involves a $K$-dimensional integral over the latent trait vector for each individual, which is difficult to evaluate numerically when $K$ is large. Second, in our setting the latent distribution is unknown and modeled by the flexible density $\pi_{\eta}(\cdot)$, adding further complexity to the optimization problem. In what follows, we propose a new estimation procedure that introduces a flow-based conditional posterior approximation, allowing the item parameters $\psi$ and the latent distribution parameters $\eta$ to be efficiently estimated together with the posterior approximation.

\section{Estimation Method}\label{sec_proposed_method}
We begin by describing the expectation-maximization (EM) algorithm for solving \eqref{eq_mmle}, a widely used approach for marginal maximum likelihood estimation in MIRT~\citep{Bock_Aitkin_1981}. The EM algorithm treats the latent traits $\theta_i$ as missing data and alternates between two steps. In the E-step, one computes the conditional distribution of $\theta_i$ given the observed response vector $Y_i$ under the current model parameters $\psi$ and $\eta$:
\begin{equation}
    p_{\psi, \eta}(\theta \mid Y_i) \; = \; \frac{p_{\psi,\eta}(Y_i, \theta_i)}{p_{\psi, \eta}(Y_i)}\label{eq_conditioning_dens}
\end{equation}
In the M-step, the model parameters are updated by maximizing the expected complete-data log-likelihood with respect to this conditional distribution. 

Directly applying the EM scheme, however, faces several challenges. First, the posterior involves the same $K$-dimensional integral that appears in the marginal likelihood, which becomes computationally expensive when $K$ is large. Second, the latent distribution $\pi_{\eta}$ is itself parameterized by a flow transformation and estimated jointly with the item parameters $\psi$, the posterior depends on both sets of parameters. This makes the EM updates difficult to implement and can also introduce identifiability concerns: without additional regularization or constraints, the location, scale, and orientation of the latent space are not uniquely determined, whereas they are typically anchored by the Gaussian specification in conventional MIRT estimation.

To address these challenges, we replace the exact posterior in the E-step with a tractable conditional density $q_{\phi}(\theta\mid Y_i)$, parameterized by $\phi$, to approximate the posterior distribution of the latent trait given the observed response vector. Specifically, we parameterize this density through a conditional flow-based sampling scheme
\begin{equation}
    \theta_i\; =\; g_{\phi}(Y_i,\epsilon_i),\qquad\epsilon_i\sim \cN(0,I_K),\label{eq_condition_flow}
\end{equation}
where $g_{\phi}$ maps each observed response vector $Y_i$ and an auxiliary Gaussian noise vector $\epsilon_i$ to the latent space. We use the notation $\epsilon_i$ for the noise in the conditional flow to distinguish it from $\varepsilon_i$ in the prior flow in \eqref{eq_raw_prior_flow}. When $g_{\phi}$ is invertible with respect to $\epsilon_i$ for each fixed $Y_i$, the induced density $q_{\phi}(\theta\mid Y_i)$ can be evaluated using the change-of-variables formula similar to \eqref{eq_change_of_var}.

Plugging the parameterized approximate posterior $q_{\phi}(\theta|Y_i)$ into the expected complete-data log-likelihood and adding its entropy term yields the following:
\begin{equation}
    \bar\cL(\psi,\eta,\phi) \; := \; \sum_{i=1}^N
    \mathbb{E}_{\theta_i\sim q_{\phi}(\cdot\mid Y_i)}\left[\log p_{\psi}(Y_i\mid \theta_i) + \log \pi_{\eta}(\theta_i) - \log q_{\phi}(\theta_i\mid Y_i)\right].
\label{eq_elbo}
\end{equation}
The first term is the expected conditional response log-likelihood, the second term is the expected logarithm of prior density, and the third term is the entropy contribution of the approximate posterior. This entropy term arises from the variational formulation and can be interpreted as an entropy regularization term that discourages overly concentrated posterior approximations. The objective $\bar\cL(\psi,\eta,\phi)$ is a lower bound on the marginal log-likelihood and becomes tight when $q_\phi(\theta_i\mid Y_i)$ coincides with the exact posterior distribution of $\theta_i$ given $Y_i$ under the model; see the discussion below. We propose to jointly estimate the item parameters $\psi$, prior-flow parameters $\eta$, and posterior-flow parameters $\phi$ by solving
\begin{equation}
    (\hat\psi,\hat\eta,\hat\phi)\;\; =\;\; \argmax_{\psi,\eta,\phi}\; \bar\cL(\psi,\eta,\phi),\label{eq_elbo_estimation}
\end{equation}
 A stochastic optimization to compute \eqref{eq_elbo_estimation} is presented in Section~\ref{sec_alg} in the following.

 One notable advantage of this formulation is that it avoids estimating a separate density $q_i(\theta)$ for each individual at each parameter update. Instead, the conditional density $q_{\phi}(\theta\mid Y_i)$ is learned jointly across all individuals by optimizing \eqref{eq_elbo}. For fixed item parameters $\psi$ and prior-flow parameters $\eta$, optimizing with respect to $\phi$ encourages $q_{\phi}(\theta\mid Y_i)$ to approximate the exact posterior distribution of $\theta_i$ given $Y_i$ under the current model. Simultaneously, this reduces the gap between the objective $\bar\cL(\psi,\eta,\phi)$ and the marginal log-likelihood. To see this, define the following 
Kullback-Leibler (KL) divergence
\begin{align*}
D_{\mathrm{KL}} \left\{ q_{\phi}(\theta\mid Y_i)\,\|\, p_{\psi,\eta}(\theta\mid Y_i) \right\} \;:= &\;\int_{\theta\in\RR^{K}}q_{\phi}(\theta\mid Y_i)\log\frac{q_{\phi}(\theta\mid Y_i)}{p_{\psi,\eta}(\theta\mid Y_i)}d\theta\\ = &\; \mathbb{E}_{\theta_i\sim q_{\phi}(\cdot\mid Y_i)}\left[\log\big\{q_{\phi}(\theta\mid Y_i)\big\} -\log\big\{p_{\psi,\eta}(\theta\mid Y_i)\big\}\right].
\end{align*}
Using \eqref{eq_conditioning_dens}, we know 
\begin{equation}\bar\cL(\psi,\eta,\phi)  + \sum_{i=1}^ND_{\mathrm{KL}} \left\{ q_{\phi}(\theta\mid Y_i)\,\|\, p_{\psi,\eta}(\theta\mid Y_i) \right\} = \sum_{i=1}^N\log\int_{\theta_i\in\RR^K}\Pr(Y_i\mid\theta_i,\psi) \pi_{\eta}(\theta_i)d\theta_i\, ,\label{eq_relate_variational}
\end{equation}where the right side is the marginal likelihood of the observed responses under parameters $(\psi,\eta)$.
For fixed $\psi$ and $\eta$, this quantity does not depend on $\phi$. Therefore, maximizing \eqref{eq_elbo} with respect to $\phi$ is equivalent to minimizing the aggregate KL divergence
$\sum_{i=1}^MD_{\mathrm{KL}} \left\{ q_{\phi}(\theta\mid Y_i)\,\|\, p_{\psi,\eta}(\theta\mid Y_i) \right\} $. Thus, optimizing $\phi$ makes the conditional density $q_{\phi}(\theta_i\mid Y_i)$ approximate the numerically intractable posterior $p_{\psi,\eta}(\theta_i\mid Y_i)$ as well as possible within the chosen flow-based family. If the chosen flow-based family is sufficiently expressive and the optimization is accurate, the KL gap can be made small and close to 0, so that optimizing $\bar\cL(\psi,\eta,\phi)$ provides a close approximation to marginal maximum likelihood estimation.

\begin{remark}\label{remark_variational}
The proposed procedure is related to the variational EM algorithm~\citep{rijmen2013fitting,cho2021gaussian}. To see the connection, we note that the variational EM framework uses the following identity: for any density $q_i(\theta_i)$,
\begin{equation}
\begin{aligned}
    \sum_{i=1}^N\log p_{\psi}(Y_i) = &\,\sum_{i=1}^N\int_{\theta_i\in\RR^K} q_i(\theta)d\theta \log \left\{\frac{p_{\psi}(Y_i,\theta_i)}{q(\theta_i)}\times \frac{q(\theta_i)}{{p_{\psi}(Y_i, \theta_i)}/{p_{\psi}(Y_i)}}\right\}\\ =&\,\sum_{i=1}^N\int_{\theta_i\in\RR^K} q_i(\theta)\log \frac{p_{\psi}(Y_i,\theta_i)}{q(\theta_i)}d\theta + \sum_{i=1}^ND_{\rm KL}\left\{q_i(\theta)\,\|\,p_{\psi}(\theta\mid Y_i)\right\},
\end{aligned} \label{eq_decompose_ml_i}
\end{equation}
 where $p_{\psi}(\theta \mid Y_i) = \frac{p_{\psi}(Y_i, \theta_i)}{p_{\psi}(Y_i)}$ is the posterior density of $\theta_i$ given $Y_i$ under the model. Since the KL divergence is nonnegative, the first term on the right-hand side of \eqref{eq_decompose_ml_i} is always no larger than the marginal log-likelihood $\sum_{i=1}^N\log p_{\psi}(Y_i) $. This term is therefore an evidence lower bound (ELBO) on the marginal log-likelihood. 
The update of variational EM often start with choosing the density $q_i(\cdot)$  within the specified variational family to make the ELBO as tight as possible by minimizing the KL divergence. After this approximation is obtained, the model parameters are updated by maximizing the ELBO with $q(\cdot)$ fixed.

In our setting, we take $q_i(\cdot)$ to be the conditional flow density $q_{\phi}(\theta\mid Y_i)$, as in \eqref{eq_relate_variational}. The parameter $\phi$ is optimized so that $q_{\phi}(\theta\mid Y_i)$ approximates the exact posterior $p_{\psi,\eta}(\theta\mid Y_i)$, while the item parameters $\psi$ and the latent distribution parameters $\eta$ are updated by maximizing the same ELBO. Unlike standard variational EM, which may require updating a separate variational distribution for each individual at each iteration, our approach uses a shared parameterized conditional density $q_{\phi}(\theta\mid Y_i)$ for all $i\in[N]$, with individual-specific dependence entering through the observed response vector $Y_i$. This construction provides substantial flexibility, since the approximating family is no longer restricted to simple forms such as Gaussian distributions. This flexibility is important because misspecifying the approximating family can result in a nonvanishing KL divergence, leading to a loose lower bound on the marginal log-likelihood and potentially biased parameter estimates. In contrast, by increasing the capacity of $g_{\phi}$, the induced density $q_{\phi}(\theta\mid Y_i)$ can represent more complex, non-Gaussian posterior shapes and thereby provide a more accurate approximation to the exact posterior.
\end{remark}

\subsection{Algorithm}\label{sec_alg}

We compute \eqref{eq_elbo_estimation} numerically using an efficient stochastic first-order optimization algorithm, summarized in Algorithm~\ref{alg:proposed_training}. At each iteration $t$, the algorithm constructs a stochastic approximation to the objective $\bar\cL_t(\psi,\eta,\phi)$ using the minibatch Monte Carlo approximation, which is detailed later in Equation~\eqref{eq_mc_elbo}. This is obtained by first sampling a minibatch $\mathcal B^{(t)}\subset[N]$ and, for each respondent $i\in\mathcal B^{(t)}$, drawing $M$ Monte Carlo samples to approximate the expectation terms in
$\sum_{i=1}^N\mathbb{E}_{\theta_i\sim q_{\phi}(\cdot\mid Y_i)}[\cdot]$.
Since the conditional density $q_{\phi}(\cdot\mid Y_i)$ is induced by the conditional flow-based sampling scheme in \eqref{eq_condition_flow}, each latent trait sample is generated through a reparameterization step. In particular, we can simply draw Gaussian noise $\epsilon_{im}^{(t)}\sim\cN(0,I_K)$ and then set
$\theta_{im}^{(t)}=g_{\phi}(Y_i,\epsilon_{im}^{(t)})$. To evaluate the prior-flow term
$\sum_{i=1}^N\mathbb{E}_{\theta_i\sim q_{\phi}(\cdot\mid Y_i)}[\log \pi_{\eta}(\theta_i)]$,
where $\pi_{\eta}(\cdot)$ is parameterized by the prior flow in \eqref{eq_raw_prior_flow}, we compute the inverse transform
$\xi_{im}^{(t)}=h_{\eta}^{-1}(\theta_{im}^{(t)})$. This quantity is then used to evaluate $\log \pi_{\eta}(\theta_{im}^{(t)})$ through the change-of-variables formula, as detailed in \eqref{eq_explicit_elbo_general_3} below.

The gradient of the Monte Carlo objective with respect to the item parameters $\psi$ can be computed directly. The gradients with respect to the flow parameters $\eta$ and $\phi$ are computed by backpropagation, since the latent variables are generated from reparameterized Gaussian noise through differentiable flow transformations. We update $(\psi,\eta,\phi)$ using AdamW, equivalently by applying AdamW to the minimization objective $-\bar\cL_t(\psi,\eta,\phi)$ with weight decay applied in the decoupled form of \citet{loshchilov2017decoupled}.
{\spacingset{1.2}\begin{algorithm}[htbp]
\caption{Stochastic gradient optimization}
\label{alg:proposed_training}
\begin{algorithmic}[1]
\Require Response matrix $Y$, latent dimension $K$, minibatch size $B$, Monte Carlo size $M$, number of iterations $T$.
\State Initialize: item parameters $\psi$; prior flow parameters $\eta$; conditional flow parameters $\phi$;
\For{$t=1,2,\dots,T$}
    \State Sample a minibatch $\mathcal{B}^{(t)}\subset[N]$ with $|\mathcal{B}^{(t)}|=B$
    \For{each $i\in\mathcal{B}^{(t)}$}
        \For{$m=1,\dots,M$}
            \State Sample $\epsilon_{im}^{(t)}\sim \cN(0,I_K)$
            \State Set $\theta_{im}^{(t)}=g_{\phi}(Y_i,\epsilon_{im}^{(t)})$
            \State Set $\xi_{im}^{(t)}=h_{\eta}^{-1}(\theta_{im}^{(t)})$
        \EndFor
    \EndFor
    \State Construct the stochastic objective $\bar{\cL}_{t}(\psi,\eta,\phi)$ using \eqref{eq_mc_elbo} and the samples $\{\theta_{im}^{(t)},\xi_{im}^{(t)}\}_{i\in\mathcal{B}^{(t)},m\in[M]}$
    \State Compute the gradient $\nabla_{\psi,\eta,\phi}\bar{\cL}_{t}(\psi,\eta,\phi)$
    \State Update $(\psi,\eta,\phi)$ by applying AdamW to $-\bar{\cL}_{t}(\psi,\eta,\phi)$
\EndFor
\State \Return $(\hat\psi,\hat\eta,\hat\phi)$
\end{algorithmic}
\end{algorithm}}

Now we make the objective in \eqref{eq_elbo} explicit. Write $p_{\cN}(\cdot) = p_{\cN}(\cdot\mid 0,I_K)$ as the density of $\cN(0,I_K)$ for simplicity. Recall that the conditional flow density $q_{\phi}(\theta_i\mid Y_i)$ is induced by
\begin{equation*}
    \theta_i = g_{\phi}(Y_i,\epsilon_i), \qquad
    \epsilon_i\sim \cN(0,I_K),
\end{equation*}
where $g_{\phi}$ is invertible with respect to $\epsilon_i$ for each fixed $Y_i$. Then for any function $f(\theta)$, the expectation $\mathbb{E}_{q_{\phi}(\theta|Y_i)}[f(\theta)]$ can be written as 
\begin{equation*}
    \mathbb{E}_{\theta\sim q_{\phi}(\theta|Y_i)}[f(\theta)] = \mathbb{E}_{\epsilon\sim p_{\cN}(\cdot)}[f(g_{\phi}(Y_i,\epsilon))].
\end{equation*}
Subsequently, \eqref{eq_elbo} can be expressed as 
\begin{align}
    \bar\cL(\psi,\eta,\phi)  = &\,
    \sum_{i=1}^N\mathbb{E}_{\epsilon_i\sim \cN(0,I_K)}\Big[\log p_{\psi}\big(Y_i\mid g_{\phi}(Y_i,\epsilon_i)\big)\Big]\label{eq_explicit_elbo_general_1}\\&\, + \sum_{i=1}^N\mathbb{E}_{\epsilon_i\sim \cN(0,I_K)}\Big[\log \pi_{\eta}\big(g_{\phi}(Y_i,\epsilon_i)\big) - \log q_{\phi}\big(g_{\phi}(Y_i,\epsilon_i)\mid Y_i\big)\Big].
\label{eq_explicit_elbo_general_2}
\end{align}
For \eqref{eq_explicit_elbo_general_1}, we note that under the M2PL model, there is
\begin{align*}
    \log p_{\psi}\left(Y_i\mid g_{\phi}(Y_i,\epsilon_i)\right) = \sum_{j=1}^J \Big[&\,Y_{ij} \log\sigma\left\{\alpha_j^\T g_{\phi}(Y_i,\epsilon_i)-b_j \right\} \\&\,+ (1-Y_{ij}) \log \big\{1-\sigma\big(\alpha_j^\T g_{\phi}(Y_i,\epsilon_i)-b_j\big)\big\}\Big].
\end{align*}
Then \eqref{eq_explicit_elbo_general_1} can be calculated as
\begin{equation}
\begin{aligned}
    \eqref{eq_explicit_elbo_general_1} = 
    \sum_{i=1}^N\mathbb{E}_{\epsilon_i\sim \cN(0,I_K)}\sum_{j=1}^J\Big\{&Y_{ij}\log\sigma\left(\alpha_j^\T g_{\phi}(Y_i,\epsilon_i)-b_j\right)\\
        &+(1-Y_{ij})\log\left[1-\sigma\left(\alpha_j^\T g_{\phi}(Y_i,\epsilon_i)-b_j\right)\right]\bigg\}
\end{aligned}
\label{eq_explicit_elbo}
\end{equation}
Next, for \eqref{eq_explicit_elbo_general_2}, we note that the latent prior density $\pi_{\eta}(\cdot)$ is also induced by the prior flow \eqref{eq_raw_prior_flow}.
When evaluating $\pi_{\eta}(\theta_i)$ at a latent draw $\theta_i=g_{\phi}(Y_i,\epsilon_i)$, we define
\begin{equation*}
    \xi_i = h_{\eta}^{-1}(\theta_i) = h_{\eta}^{-1}\{g_{\phi}(Y_i,\epsilon_i)\}.
\end{equation*}
Thus, by the change-of-variables formula, we can write
\begin{equation*}
    \log \pi_{\eta}(\theta_i) = \log p_{\cN}(\xi_i) - \log \Big| \det \frac{\partial h_{\eta}(\xi_i)}{\partial \xi_i}\Big|.
\end{equation*}
For $\log q_{\phi}(\theta_i\mid Y_i)$, we have
\begin{equation*}
    \log q_{\phi}(\theta_i\mid Y_i) =\log p_{\cN}(\epsilon_i) - \log \Big|\det\frac{\partial g_{\phi}(Y_i,\epsilon_i)} {\partial \epsilon_i} \Big|.
\end{equation*}
Then \eqref{eq_explicit_elbo_general_2} becomes
\begin{equation}
\begin{aligned}
    \eqref{eq_explicit_elbo_general_2}  =
    \sum_{i=1}^N\mathbb{E}_{\epsilon_i\sim \cN(0,I_K)}\Big[\log p_{\cN}(\xi_i) - \log\Big|\det\frac{\partial h_{\eta}(\xi_i)}{\partial \xi_i}\Big|-\log p_{\cN}(\epsilon_i) + \log \Big|\det \frac{\partial g_{\phi}(Y_i,\epsilon_i)}{\partial \epsilon_i}\Big|\Big].\end{aligned}
\label{eq_explicit_elbo_general_3}
\end{equation}
Now we have transformed the two components of $\bar\cL(\psi,\eta,\phi) $ in terms of expectation with respect to the tractable distribution $\cN(0,I_K)$, which can be efficiently approximated by Monte Carlo samples. Specifically, at the $t$-th step, as we have selected the minibatch $\mathcal{B}^t \subset[N]$, we draw $\epsilon_{im}\sim \cN(0,I_K)$ for $m=1,\dots,M$ and $i\in\mathcal{B}^t$. Let $ \theta_{im} = g_{\phi}(Y_i,\epsilon_{im})$ and $\xi_{im} = h_{\eta}^{-1}(\theta_{im})$. Combine \eqref{eq_explicit_elbo_general_1}--\eqref{eq_explicit_elbo_general_3} to yield a Monte Carlo approximation to \(\bar\cL(\psi,\eta,\phi)\) is
\begin{equation}
\begin{aligned}
    \bar\cL_t(\psi,\eta,\phi)\approx \sum_{i\in \mathcal{B}^t}\frac{1}{B}\sum_{m=1}^M\Bigg[&\sum_{j=1}^J\bigg\{Y_{ij}\log\sigma\left(\alpha_j^\T \theta_{im}-b_j\right)\\&\qquad\qquad+(1-Y_{ij})\log\left[1-\sigma\left(\alpha_j^\T \theta_{im}-b_j\right)\right]\bigg\}\\
    & \, -\frac{1}{2}\|\xi_{im}\|^2-\log\left|\det\frac{\partial h_{\eta}(\xi)}{\partial \xi}\right|_{\xi=\xi_{im}} \\
    &\, +\frac{1}{2}\|\epsilon_{im}\|^2 + \log\left|\det\frac{\partial g_{\phi}(Y_i,\epsilon)}{\partial \epsilon}\right|_{\epsilon=\epsilon_{im}}\Bigg].
\end{aligned}
\label{eq_mc_elbo}
\end{equation}
Here, we cancel the constants in $\log p_{\cN}(\xi_{im}) - \log p_{\cN}(\epsilon_{im})$. The resulting expression is the stochastic objective constructed in Line~11 of Algorithm~\ref{alg:proposed_training}.

\section{Simulation Studies}\label{sec_simu}
We conduct simulation studies to evaluate the performance of the proposed method under several latent distribution designs. We compare the estimation accuracy of both item parameters and latent traits against two widely used estimation methods: Monte Carlo EM (MCEM) and the Metropolis-Hastings Robbins-Monro (MHRM) algorithm. These benchmark methods are implemented using the \texttt{mirt} package in R~\citep{chalmers2012mirt}.

We independently generate the latent traits $\{\theta_i\}_{i\in[N]}$. For the latent distribution, we consider the following three cases:
\begin{itemize}
    \item {\bf Gaussian}: a multivariate normal distribution $\cN(0,\Sigma)$.

    \item {\bf Mixture of Gaussians}: a four-component Gaussian mixture distribution,
    \begin{equation*}
        \theta_i \sim \sum_{m=1}^{4} \pi_m \cN(\mu_m,\Sigma_m),
    \end{equation*}
    with mixing weights $\pi=(1/8,1/8,3/8,3/8)$. The component means are placed at distinct vertices in the first two latent dimensions, while the remaining coordinates are centered at zero. Each covariance matrix is set to $\Sigma_m=\sigma^2 I_K$ with $\sigma=0.3$.

    \item {\bf Student-$t$}: a multivariate Student-$t$ distribution generated as follows. We first sample $x_i \sim \cN(0,\Sigma_0)$
    where $\Sigma_0$ is a covariance matrix whose diagonal elements are 1 and whose off-diagonal elements are drawn from uniform distribution \(\mathrm{Unif}(0.5,0.7)\). We then sample $w_i \sim \chi^2_{\nu}/\nu$
    and set
    \[
        \theta_i = \frac{x_i}{\sqrt{w_i}},
    \]
    which gives a multivariate Student-$t$ distribution with $\nu=5$ degrees of freedom.
\end{itemize}

These three designs represent a standard Gaussian latent distribution, a multimodal and potentially asymmetric latent distribution, and a heavy-tailed latent distribution, respectively. For identification, after generating $\{\theta_i\}_{i\in[N]}$ from one of the above distributions, we standardize the latent traits coordinate-wise so that each dimension has sample mean zero and sample variance one, while allowing correlations across latent dimensions. This standardization does not remove the non-Gaussian features of the latent distribution, such as multimodality or heavy tails, but places the latent traits on an identifiable scale for estimation.

In addition to varying the latent distribution, we consider the number of items $J\in\{20,50\}$ and the latent dimension $K\in\{2,6,10\}$. The sample size is fixed at $N=2000$ in the simulation study. We generate the item parameters as follows. For each discrimination vector $\alpha_j$, $j\in[J]$, the $k$th component $\alpha_{jk}$ is independently generated by
\[
\alpha_{jk}=
\begin{cases} U_{jk}, & Q_{jk}=1,\\
0, & Q_{jk}=0,
\end{cases} \qquad
U_{jk}\overset{\text{i.i.d.}}{\sim}\mathrm{Unif}(1,2).
\]
Here, \(Q=(Q_{jk})_{J\times K} \in \{0,1\}^{J\times K}\) is a binary indicator matrix that specifies the sparsity pattern in the discrimination parameters. The first \(K\) items are assigned as anchor items, so that the first \(K\) rows of \(Q\) form an identity block, namely \(Q_{1:K,\,:}=I_K\). For the remaining items, the nonzero positions are assigned cyclically across the \(K\) latent dimensions. The sparsity level depends on the latent dimension: each non-anchor item has one nonzero entry when \(K=2\), two nonzero entries when \(K=6\), and three nonzero entries when \(K=10\). A detailed description of the resulting \(Q\)-matrix pattern is provided in Appendix~\ref{appendix_11}.
The difficulty parameters are independently generated as $b_j\overset{\text{i.i.d.}}{\sim}N(0,1)$.

For the latent prior flow \(h_\eta\), we use two architectural specifications depending on the complexity of the latent distribution. For complex non-Gaussian latent distributions, we use a neural spline flow~\citep{durkan2019neural} with 6 layers, hidden dimension 256, 16 spline bins, and tail bound 8.0. For latent distributions close to Gaussian, we use a simpler neural spline flow with 1 layer, hidden dimension 64, 4 spline bins, and tail bound 6.0. 
For the sampling transformation \(g_\phi(\cdot\mid Y_i)\) for conditional density $q_{\phi}(\cdot|Y_i)$, we use a neural network with four stochastic layers, and hidden dimensions 100 at each layer. 
The model is optimized using AdamW with minibatch size 512, learning rate \(4\times 10^{-4}\), and weight decay \(2\times 10^{-5}\). The optimization is terminated when the relative change in the training loss is smaller than \(10^{-4}\). 

We evaluate recovery of the item parameters using bias and root mean squared error (RMSE). Each simulation setting is repeated 50 times, and the reported results are summarized across these repetitions. To assess the quality of the recovered latent traits, we report the bias and RMSE of the posterior modes obtained from the proposed posterior approximation, treating them as point estimates.

\subsection{Results under Exploratory Analysis}\label{sec_simu_1}

We first consider the exploratory setting, in which the sparsity pattern in the discrimination parameters is unknown~\citep{fabrigar2012exploratory}. In this setting, we first estimate the model with the proposed method, and then apply a post-hoc rotation to recover the sparse structure in the discrimination parameters~\citep{browne2001overview}. To allow the latent traits to be correlated, we use promax, a commonly used oblique rotation method.

{\spacingset{1.2}
\begin{figure}[!htbp]
    \centering
    
    \includegraphics[width=0.3\textwidth]{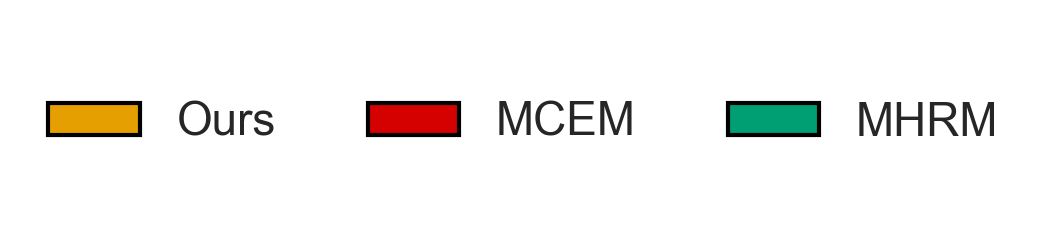}
    \vspace{0.01em}
    \begin{subfigure}[b]{0.95\textwidth}
        \centering
        \includegraphics[width=\textwidth]{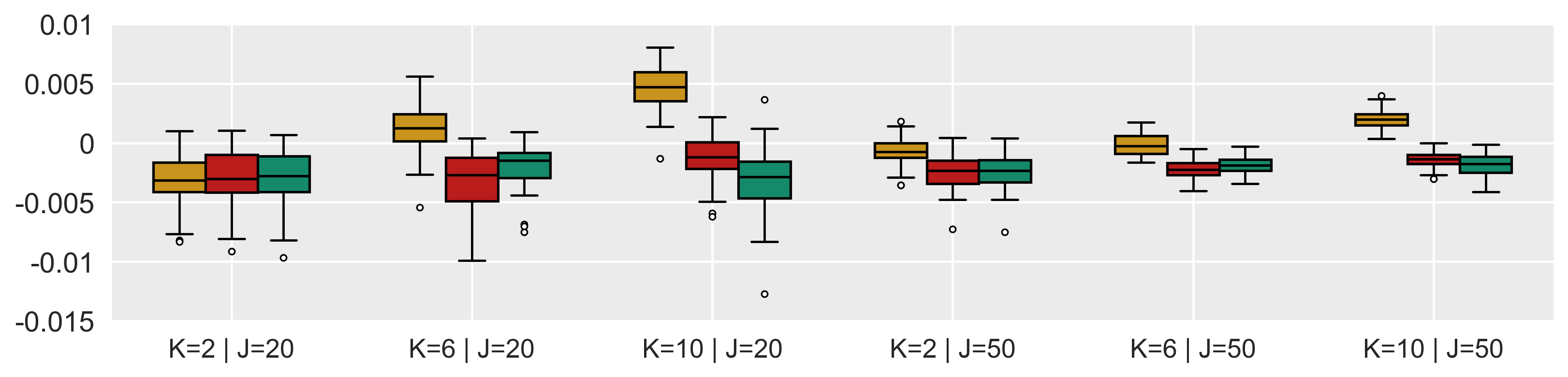}
        \caption{Bias for $\alpha$}
        \label{fig:a_bias_normal_exp}
    \end{subfigure}

    \vspace{0.1em}

    \begin{subfigure}[b]{0.95\textwidth}
        \centering
        \includegraphics[width=\textwidth]{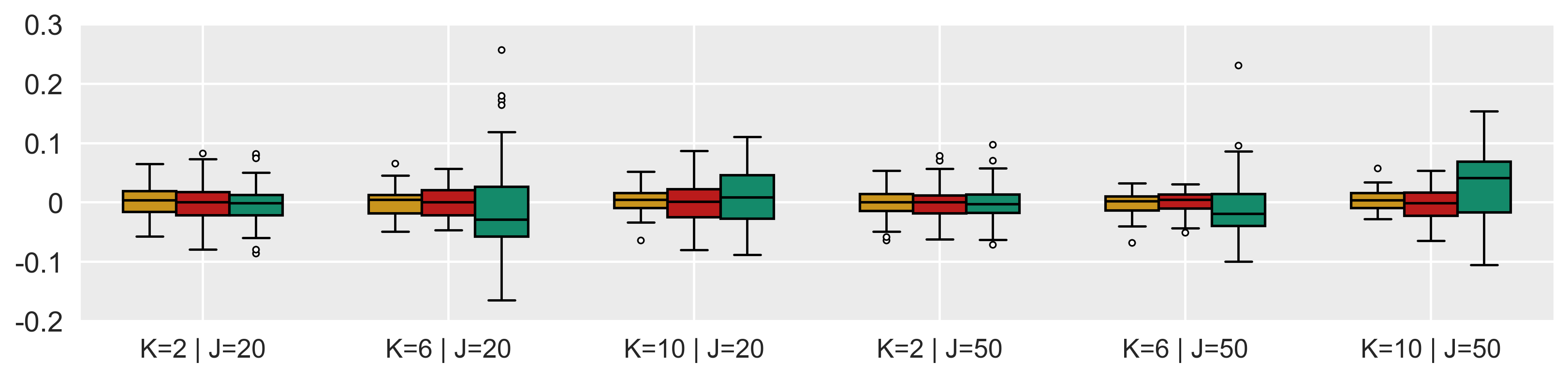}
        \caption{Bias for $b$}
        \label{fig:b_bias_normal_exp}
    \end{subfigure}

    \vspace{0.1em}

    \begin{subfigure}[b]{0.95\textwidth}
        \centering
        \includegraphics[width=\textwidth]{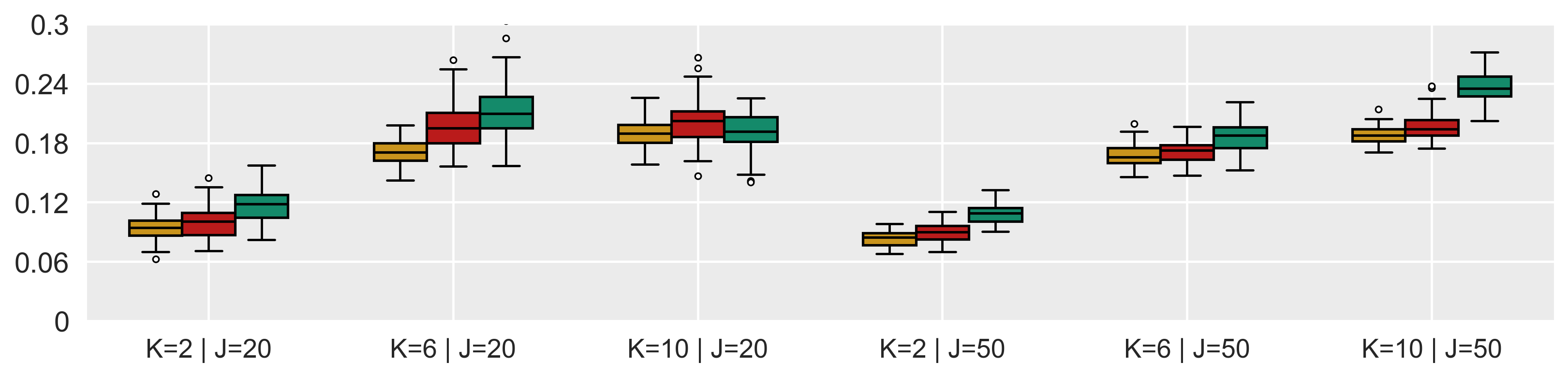}
        \caption{RMSE for $\alpha$}
        \label{fig:a_rmse_normal_exp}
    \end{subfigure}

    \vspace{0.1em}

    \begin{subfigure}[b]{0.95\textwidth}
        \centering
        \includegraphics[width=\textwidth]{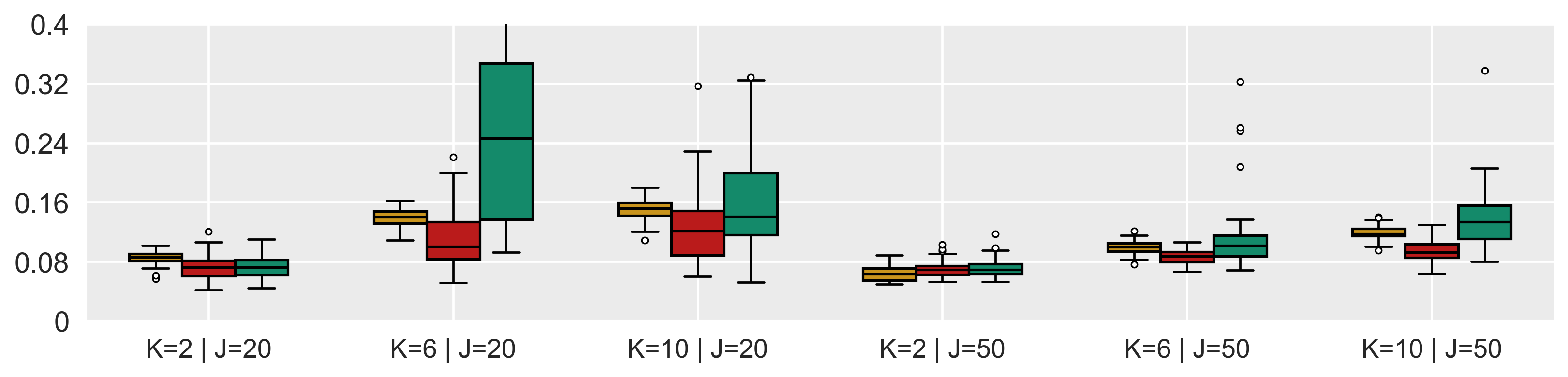}
        \caption{RMSE for $b$}
        \label{fig:b_rmse_normal_exp}
    \end{subfigure}

    \caption{Item parameter recovery for M2PL models under a Gaussian latent distribution in the exploratory setting, comparing MCEM, MHRM, and the proposed method.}
    \label{fig:normal_exp}
\end{figure}

\begin{figure}[!htbp]
    \centering
    
    \includegraphics[width=0.3\textwidth]{legend.png}
    \vspace{0.01em}
    \begin{subfigure}[b]{0.95\textwidth}
        \centering
        \includegraphics[width=\textwidth]{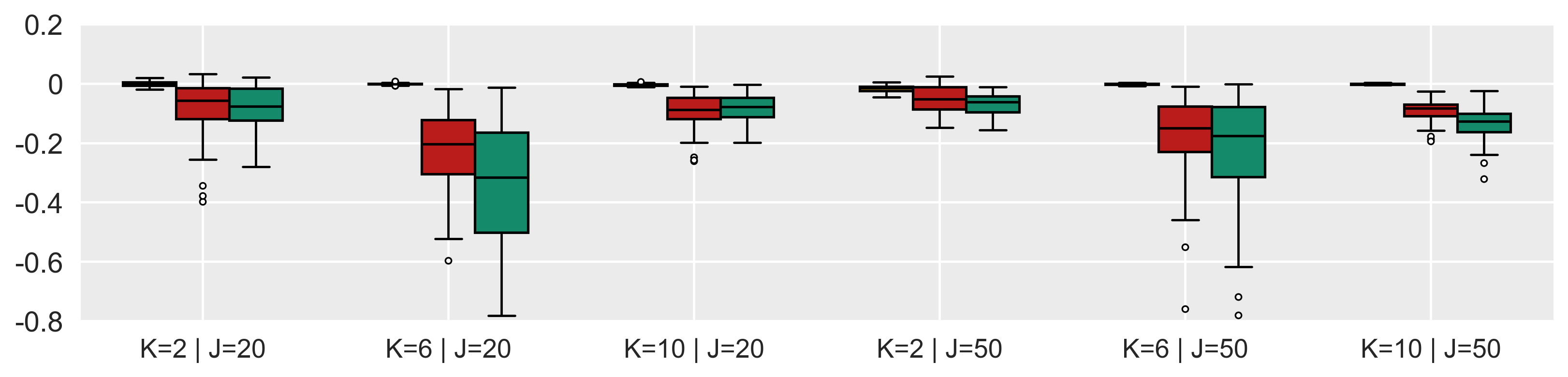}
        \caption{Bias for $\alpha$}
        \label{fig:a_bias_mog_exp}
    \end{subfigure}

    \vspace{0.1em}

    \begin{subfigure}[b]{0.95\textwidth}
        \centering
        \includegraphics[width=\textwidth]{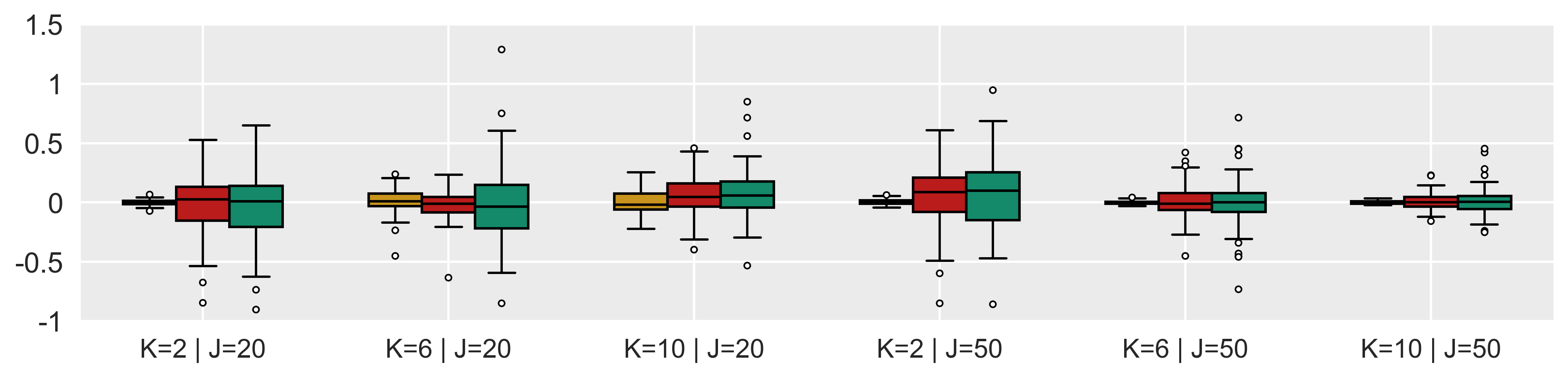}
        \caption{Bias for $b$}
        \label{fig:b_bias_mog_exp}
    \end{subfigure}

    \vspace{0.1em}

    \begin{subfigure}[b]{0.95\textwidth}
        \centering
        \includegraphics[width=\textwidth]{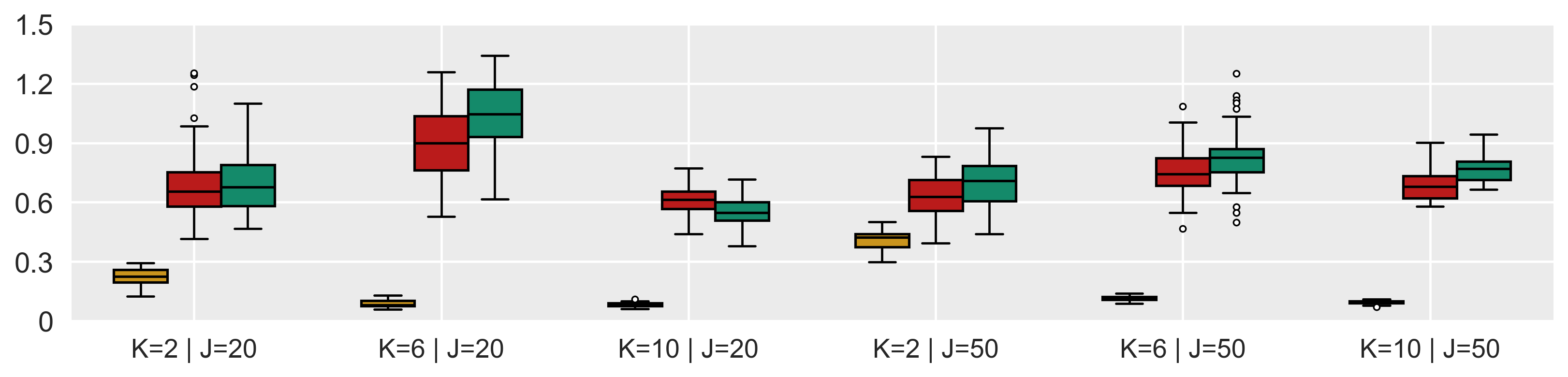}
        \caption{RMSE for $\alpha$}
        \label{fig:a_rmse_mog_exp}
    \end{subfigure}

    \vspace{0.1em}

    \begin{subfigure}[b]{0.95\textwidth}
        \centering
        \includegraphics[width=\textwidth]{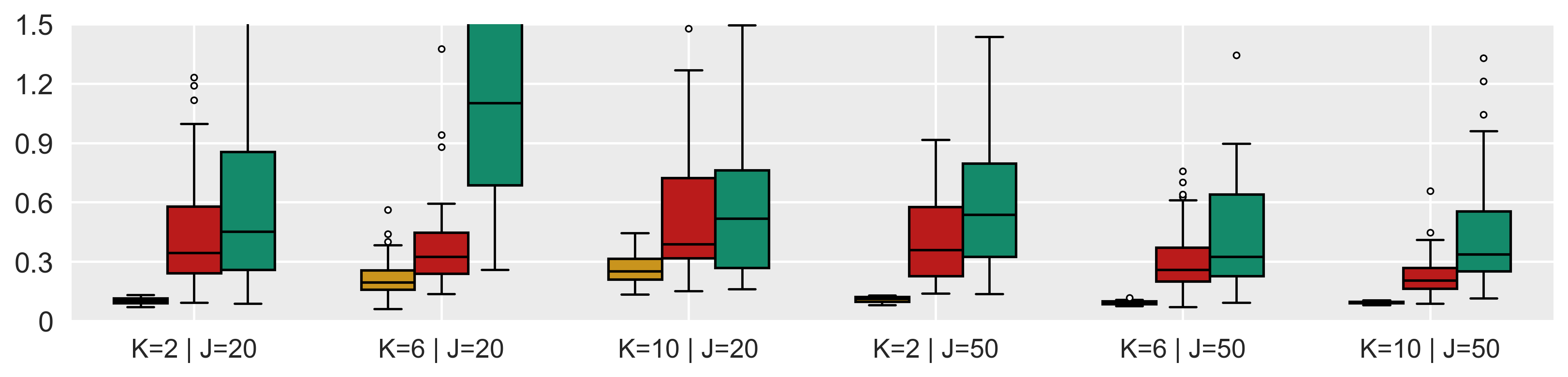}
        \caption{RMSE for $b$}
        \label{fig:b_rmse_mog_exp}
    \end{subfigure}

    \caption{Item parameter recovery for M2PL models under a mixture Gaussian latent distribution in the exploratory setting, comparing MCEM, MHRM, and the proposed method.}
    \label{fig:mog_exp}
\end{figure}

\begin{figure}[!htbp]
    \centering
    
    \includegraphics[width=0.3\textwidth]{legend.png}
    \vspace{0.01em}
    \begin{subfigure}[b]{0.95\textwidth}
        \centering
        \includegraphics[width=\textwidth]{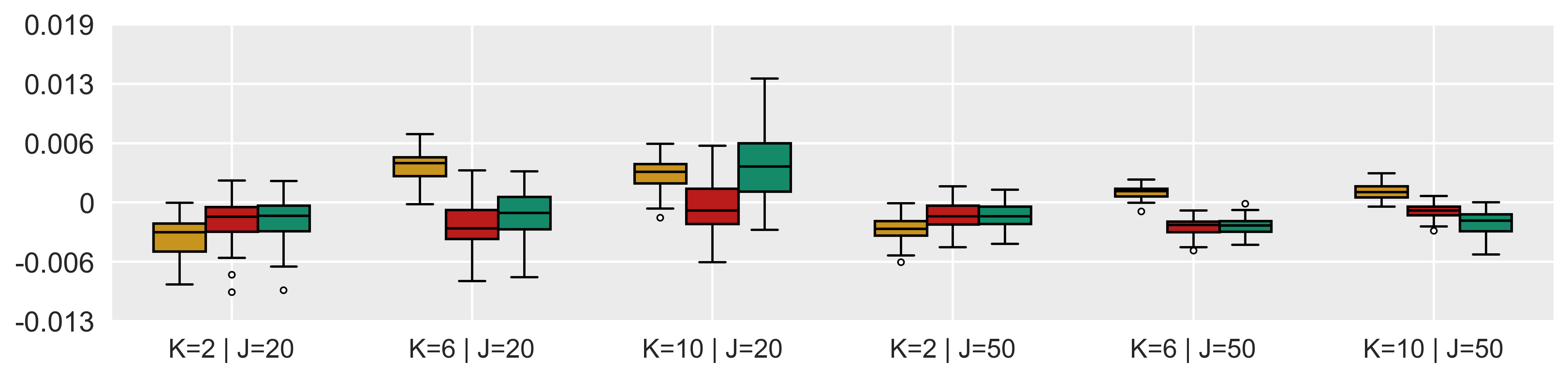}
        \caption{Bias for $\alpha$}
        \label{fig:a_bias_t_exp}
    \end{subfigure}

    \vspace{0.1em}

    \begin{subfigure}[b]{0.95\textwidth}
        \centering
        \includegraphics[width=\textwidth]{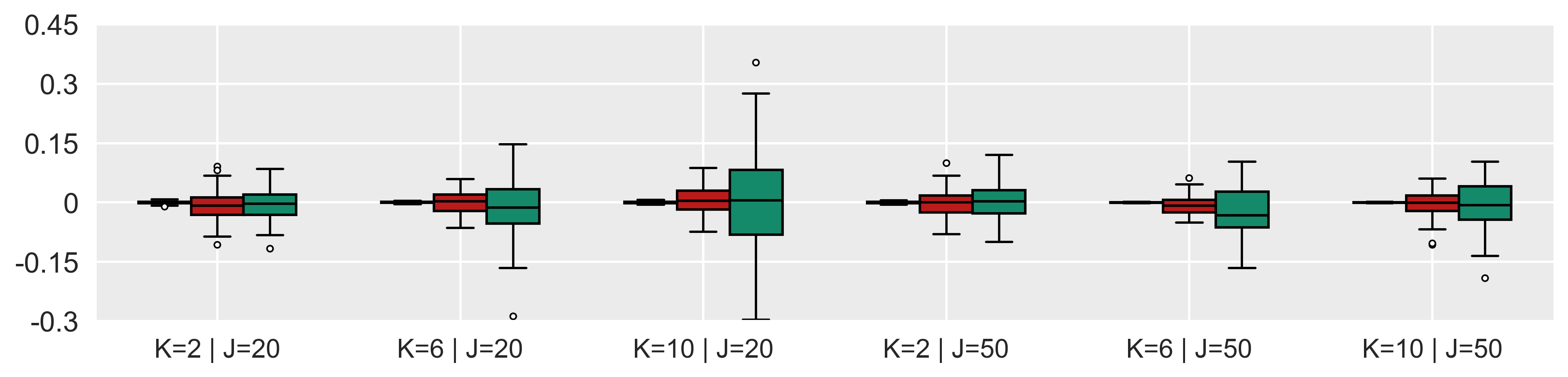}
        \caption{Bias for $b$}
        \label{fig:b_bias_t_exp}
    \end{subfigure}

    \vspace{0.1em}

    \begin{subfigure}[b]{0.95\textwidth}
        \centering
        \includegraphics[width=\textwidth]{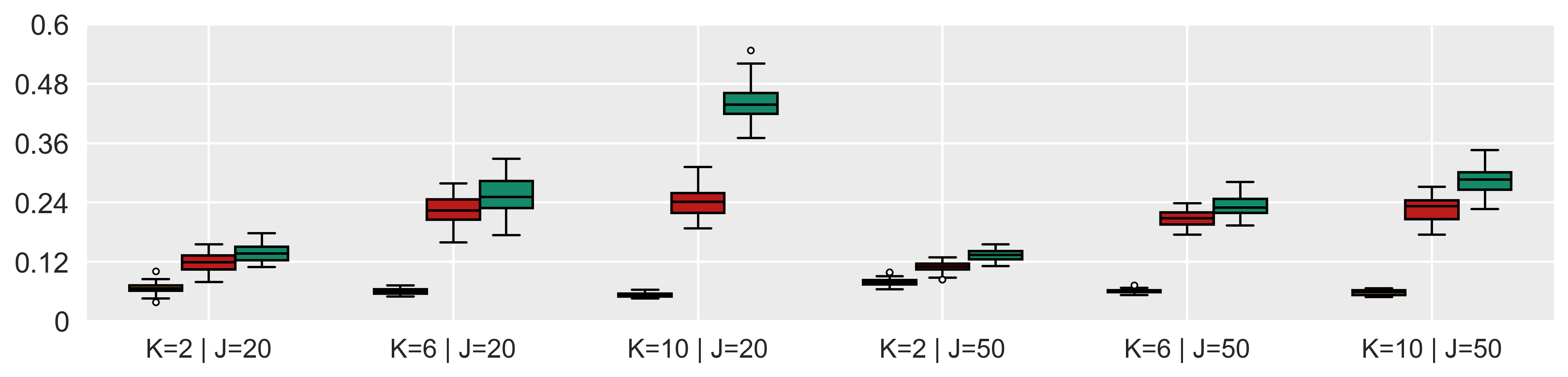}
        \caption{RMSE for $\alpha$}
        \label{fig:a_rmse_t_exp}
    \end{subfigure}

    \vspace{0.1em}

    \begin{subfigure}[b]{0.95\textwidth}
        \centering
        \includegraphics[width=\textwidth]{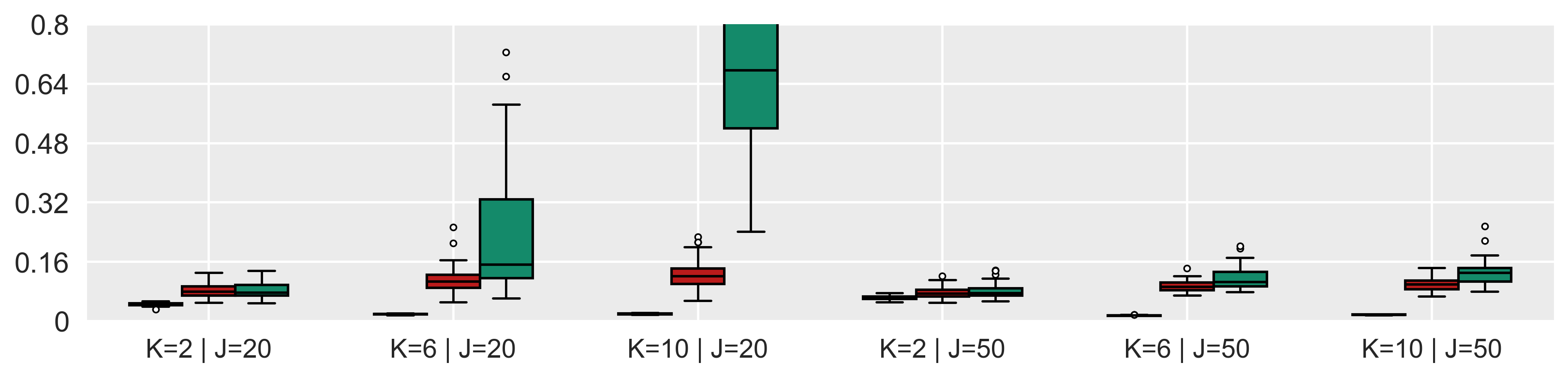}
        \caption{RMSE for $b$}
        \label{fig:b_rmse_t_exp}
    \end{subfigure}

    \caption{Item parameter recovery for M2PL models under a $t$ distribution in the exploratory setting, comparing MCEM, MHRM, and the proposed method.}
    
    \label{fig:t_exp}
\end{figure}
}


{\spacingset{1.2} \begin{figure}[H]
    \centering

    \includegraphics[width=0.25\linewidth]{legend.png}
    \vspace{0.5em}
        
\begin{subfigure}[t]{0.95\linewidth}
            \centering
            \includegraphics[width=\linewidth]{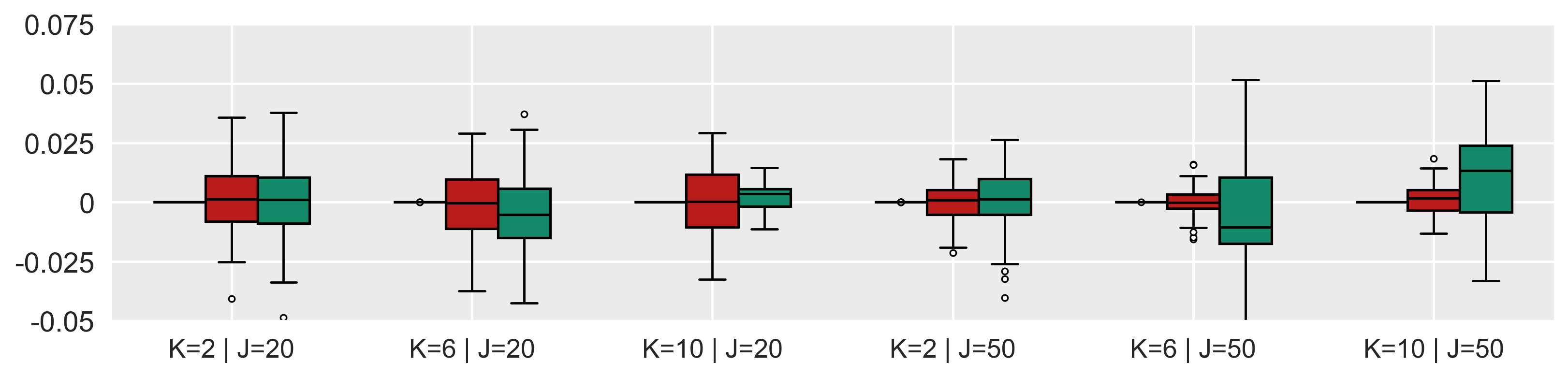}
            \caption{Gaussian}
            \label{fig:theta_bias_normal_exp}
        \end{subfigure}
        \vspace{0.5em}

        \begin{subfigure}[t]{0.95\linewidth}
            \centering
            \includegraphics[width=\linewidth]{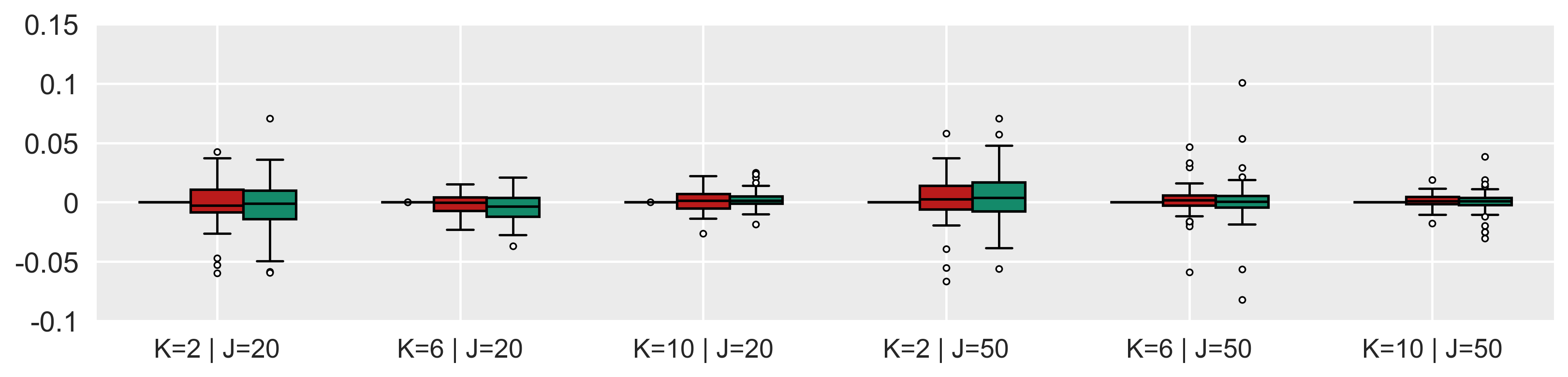}
            \caption{Mixture of Gaussian}
            \label{fig:theta_bias_mog_exp}
        \end{subfigure}

        \vspace{0.5em}

        \begin{subfigure}[t]{0.95\linewidth}
            \centering
            \includegraphics[width=\linewidth]{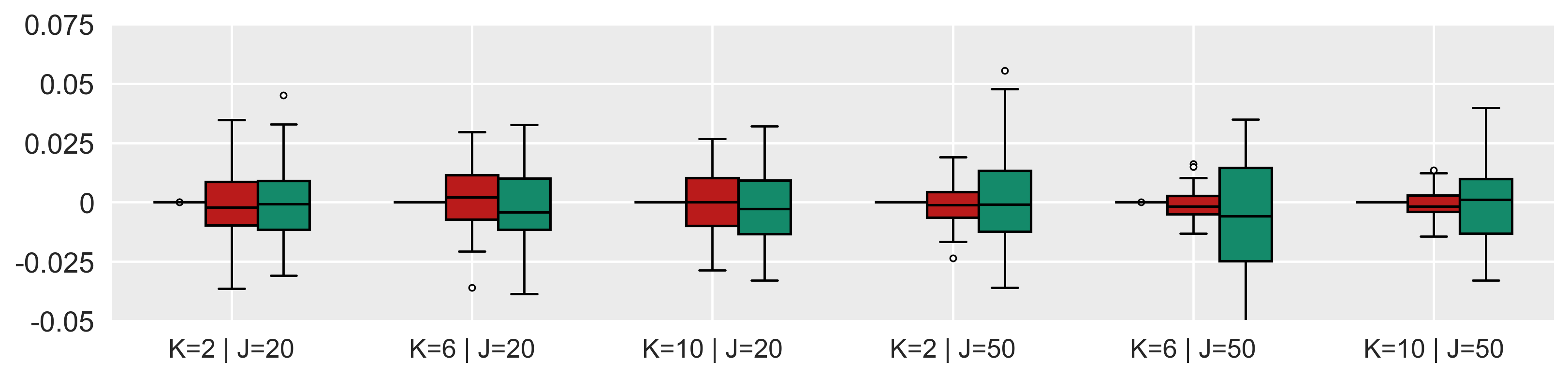}
            \caption{Student-\(t\)}
            \label{fig:theta_bias_t_exp}
        \end{subfigure}

    \caption{Bias for latent trait recovery in the exploratory setting. Rows correspond to Gaussian, Mixture of Gaussian, and Student-\(t\) latent distributions.}
    \label{fig:appendix_theta_exp_bias}
\end{figure}} 
We compare the proposed method with MCEM and MHRM. The item parameter recovery results, measured by bias and RMSE, are presented in Figures~\ref{fig:normal_exp}, \ref{fig:mog_exp}, and \ref{fig:t_exp} for the Gaussian, mixture Gaussian, and Student-$t$ latent distributions, respectively. 
From Figure~\ref{fig:normal_exp}, we can see that the bias and RMSE of the discrimination and difficulty parameter
estimates are similar across methods, suggesting that the added flexibility of the proposed framework does not substantially degrade estimation accuracy when the Gaussian assumption holds. When the latent distribution is non-Gaussian, the advantage of the proposed method becomes more pronounced. As shown in Figure~\ref{fig:mog_exp}, the proposed method also yields smaller bias and RMSE for both the discrimination parameters $\alpha$ and the difficulty parameters $b$. Similar patterns are observed under the Student-$t$ distribution, as shown in Figure~\ref{fig:t_exp}. In this heavy-tailed setting, the proposed method improves both latent distribution recovery and item parameter estimation. The performance of MCEM and MHRM is particularly problematic in the high-dimensional case with $K=10$, whereas the proposed method remains stable and accurate.

{\spacingset{1.2}\begin{figure}
    \centering

    \includegraphics[width=0.25\linewidth]{legend.png}
    \vspace{0.5em}

       \begin{subfigure}[t]{0.95\linewidth}
            \centering
            \includegraphics[width=\linewidth]{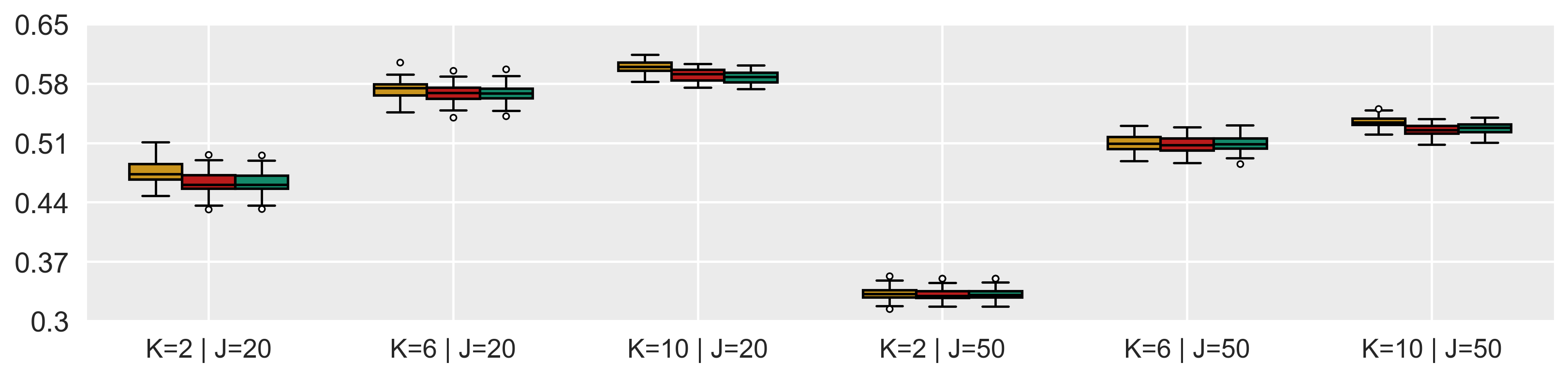}
            \caption{Gaussian}
            \label{fig:theta_rmse_normal_exp}
        \end{subfigure}

        \vspace{0.5em}

 \begin{subfigure}[t]{0.95\linewidth}
            \centering
            \includegraphics[width=\linewidth]{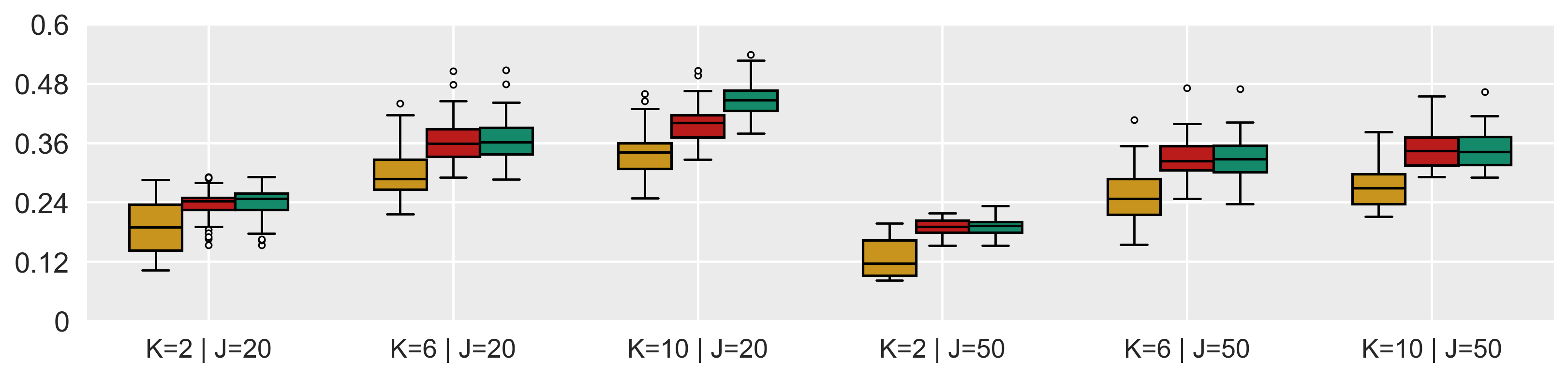}
            \caption{Mixture of Gaussian}
            \label{fig:theta_rmse_mog_exp}
        \end{subfigure}

        \vspace{0.5em}

        \begin{subfigure}[t]{0.95\linewidth}
            \centering
            \includegraphics[width=\linewidth]{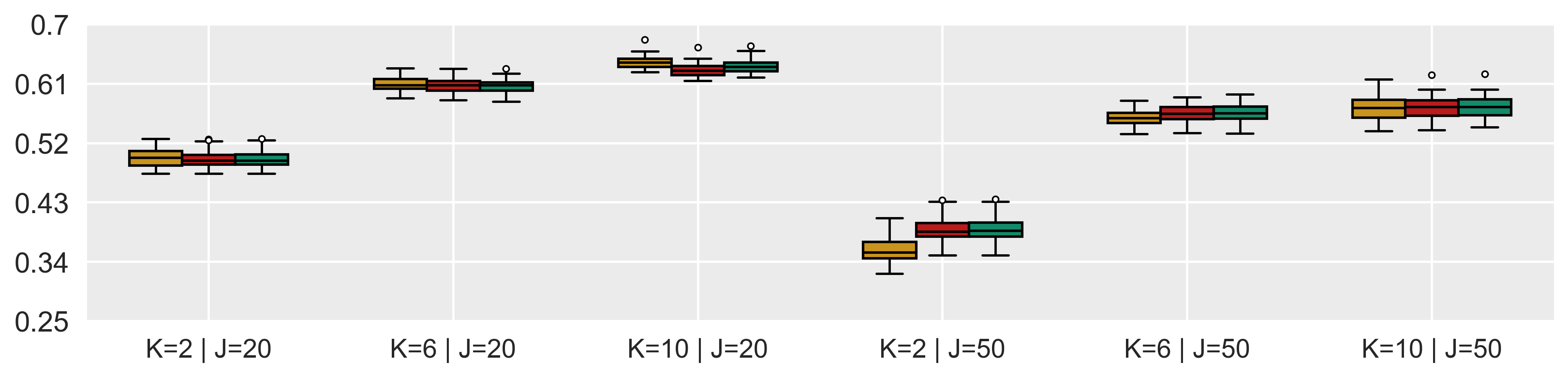}
            \caption{Student-\(t\)}
            \label{fig:theta_rmse_t_exp}
        \end{subfigure}

    \caption{RMSE for latent trait recovery in the exploratory setting. Rows correspond to Gaussian, Mixture of Gaussian, and Student-\(t\) latent distributions.}
    \label{fig:appendix_theta_exp_rmse}
\end{figure}}

Results for the point estimates of the latent traits under different settings are presented in Figures~\ref{fig:appendix_theta_exp_bias} and~\ref{fig:appendix_theta_exp_rmse}. The proposed method yields substantially smaller bias across all settings, as shown in Figure~\ref{fig:appendix_theta_exp_bias}. When the latent traits follow a Gaussian distribution, the proposed method has slightly larger RMSE than the benchmark methods in some settings; see Figure~\ref{fig:theta_rmse_normal_exp}. This is expected because the benchmark methods correctly specify the Gaussian latent distribution, whereas the proposed method treats the latent distribution as unknown and estimates it from the data. When the latent distribution is non-Gaussian, the proposed method achieves lower RMSE for latent trait recovery across different configurations of $(J,K)$, as shown in Figures~\ref{fig:theta_rmse_mog_exp} and~\ref{fig:theta_rmse_t_exp}, except a challenging setting where $(K,J)=(10,20)$. In the multi-modal setting (Figures~\ref{fig:mog_exp} and~\ref{fig:theta_rmse_mog_exp}), the proposed method improves both latent distribution recovery and item parameter estimation.
In summary, the proposed method is competitive with MCEM and MHRM when the Gaussian latent distribution is correctly specified, and it demonstrates a significant advantage in both item parameter estimation and latent trait recovery when the latent distribution is misspecified or heavy-tailed, with particularly stable performance in high-dimensional settings.

    



    
    

\subsection{Results under Confirmatory Analysis}\label{sec_simu_2}

We next consider the confirmatory setting, where the sparsity pattern $Q$ is specified in advance and the corresponding zero constraints are imposed during estimation~\citep{brown2012confirmatory}. We conduct simulation studies under the same latent distribution designs as in Section~\ref{sec_simu_1}. The item parameter recovery results, measured by bias and RMSE, are presented in Figures~\ref{fig:normal_conf}, \ref{fig:mog_conf}, and~\ref{fig:t_conf} for the Gaussian, mixture Gaussian, and Student-$t$ latent distributions, respectively. Figures~\ref{fig:appendix_theta_conf_bias} and~\ref{fig:appendix_theta_conf_rmse} report the bias and RMSE of the latent trait recovery, where the posterior modes obtained from the proposed posterior approximation are used as point estimates.

{\spacingset{1.2}
\begin{figure}[!htbp]
    \centering
    
    \includegraphics[width=0.3\textwidth]{legend.png}
    \vspace{0.01em}
    \begin{subfigure}[b]{0.95\textwidth}
        \centering
        \includegraphics[width=\textwidth]{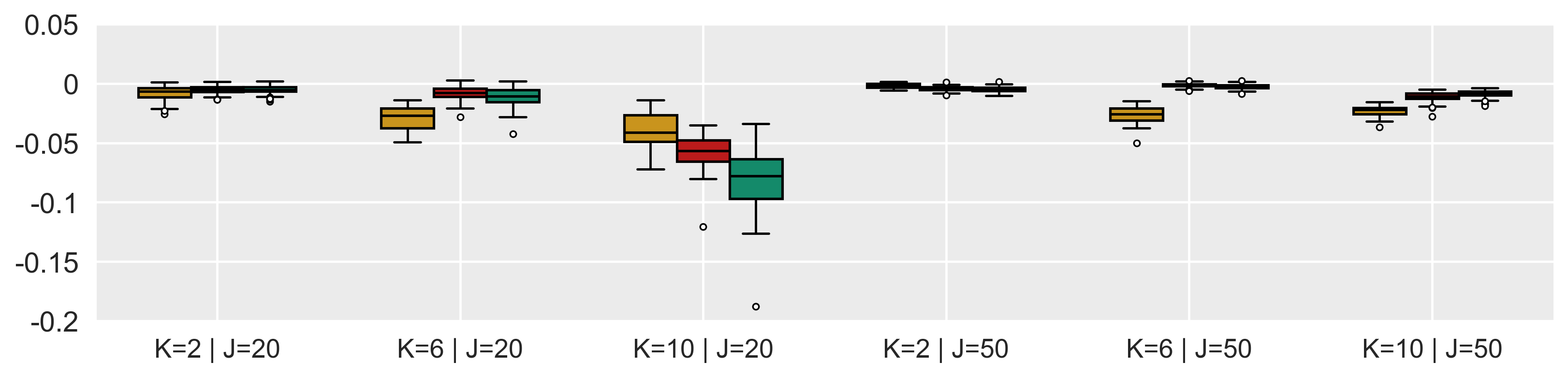}
        \caption{Bias for $\alpha$}
        \label{fig:a_bias_normal_conf}
    \end{subfigure}

    \vspace{0.1em}

    \begin{subfigure}[b]{0.95\textwidth}
        \centering
        \includegraphics[width=\textwidth]{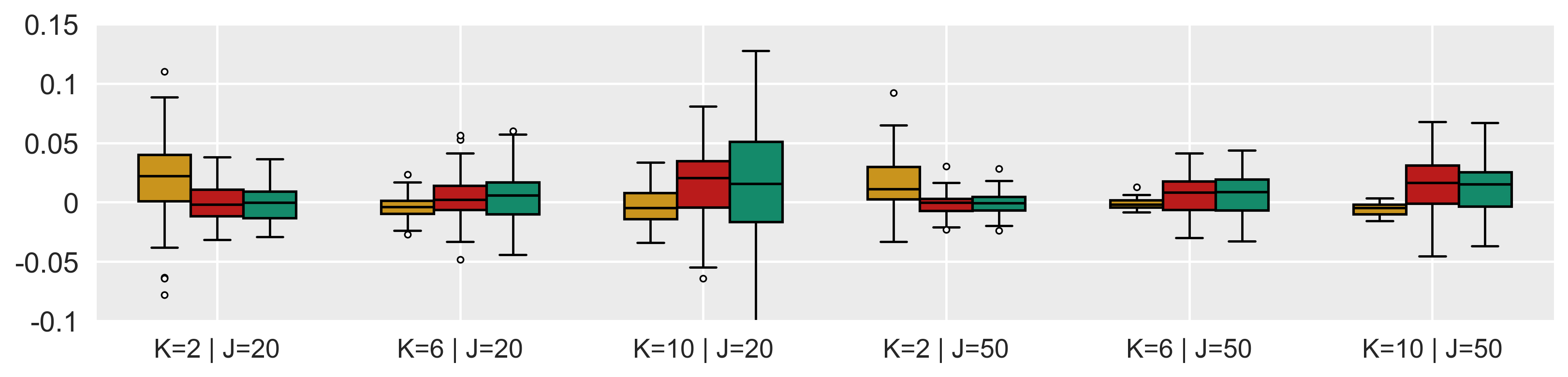}
        \caption{Bias for $b$}
        \label{fig:b_bias_normal_conf}
    \end{subfigure}

    \vspace{0.1em}

    \begin{subfigure}[b]{0.95\textwidth}
        \centering
        \includegraphics[width=\textwidth]{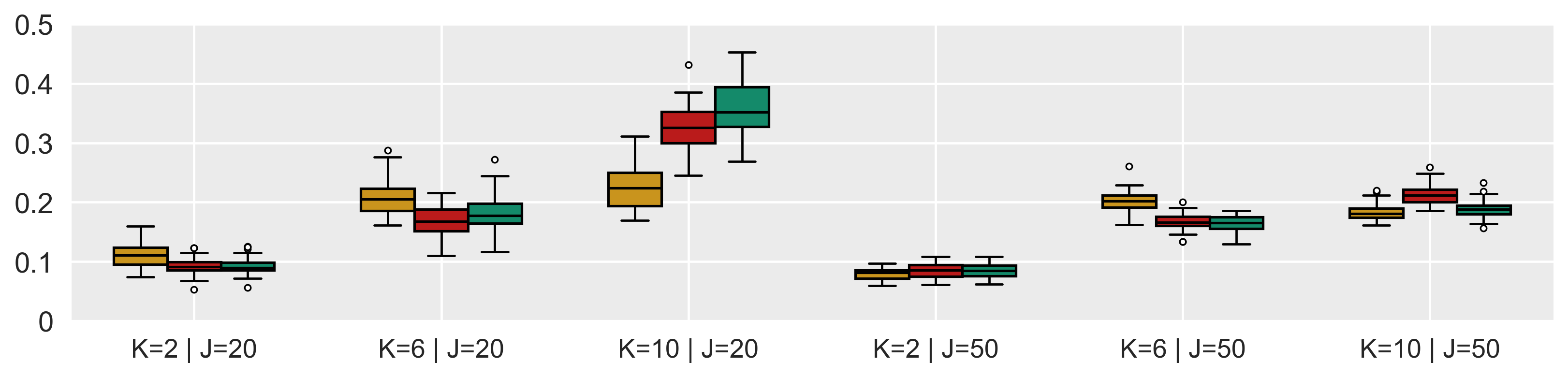}
        \caption{RMSE for $\alpha$}
        \label{fig:a_rmse_normal_conf}
    \end{subfigure}

    \vspace{0.1em}

    \begin{subfigure}[b]{0.95\textwidth}
        \centering
        \includegraphics[width=\textwidth]{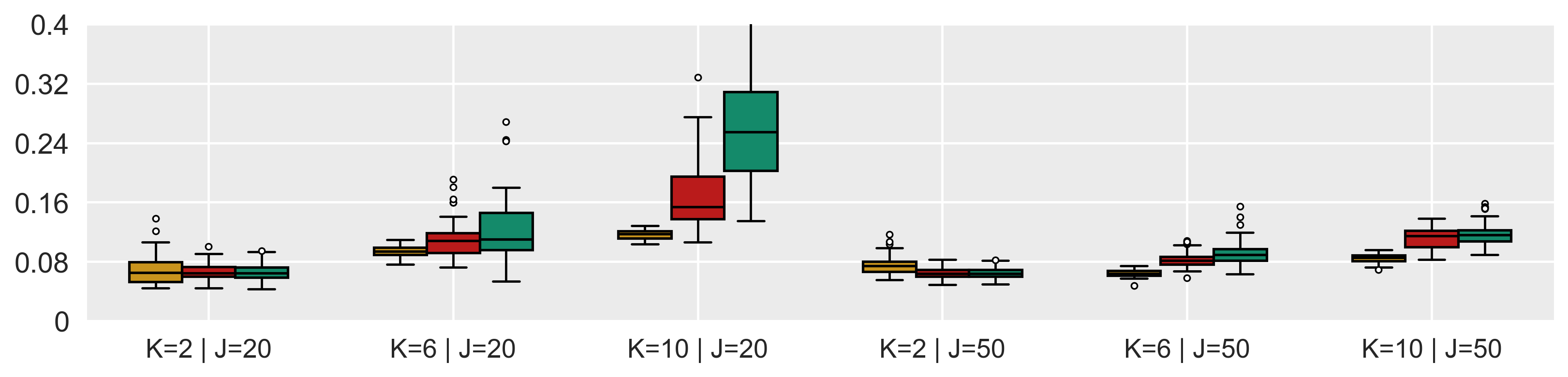}
        \caption{RMSE for $b$}
        \label{fig:b_rmse_normal_conf}
    \end{subfigure}

    \caption{Item parameter recovery for M2PL models under a Gaussian latent distribution in the confirmatory setting, comparing MCEM, MHRM, and the proposed method.}
    \label{fig:normal_conf}
\end{figure}

\begin{figure}[!htbp]
    \centering
    
    \includegraphics[width=0.3\textwidth]{legend.png}
    \vspace{0.01em}
    \begin{subfigure}[b]{0.95\textwidth}
        \centering
        \includegraphics[width=\textwidth]{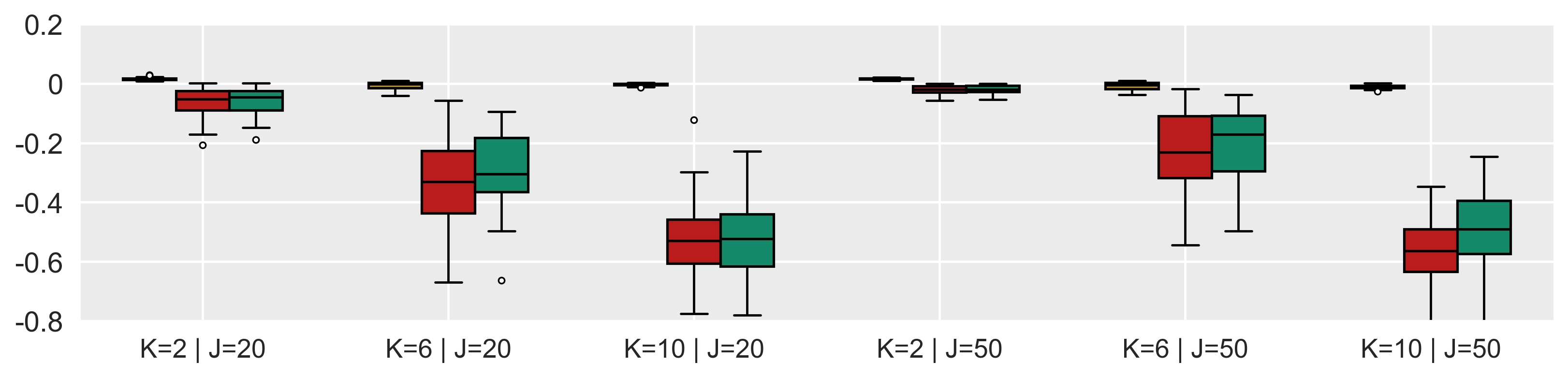}
        \caption{Bias for $\alpha$}
        \label{fig:a_bias_mog_conf}
    \end{subfigure}

    \vspace{0.1em}

    \begin{subfigure}[b]{0.95\textwidth}
        \centering
        \includegraphics[width=\textwidth]{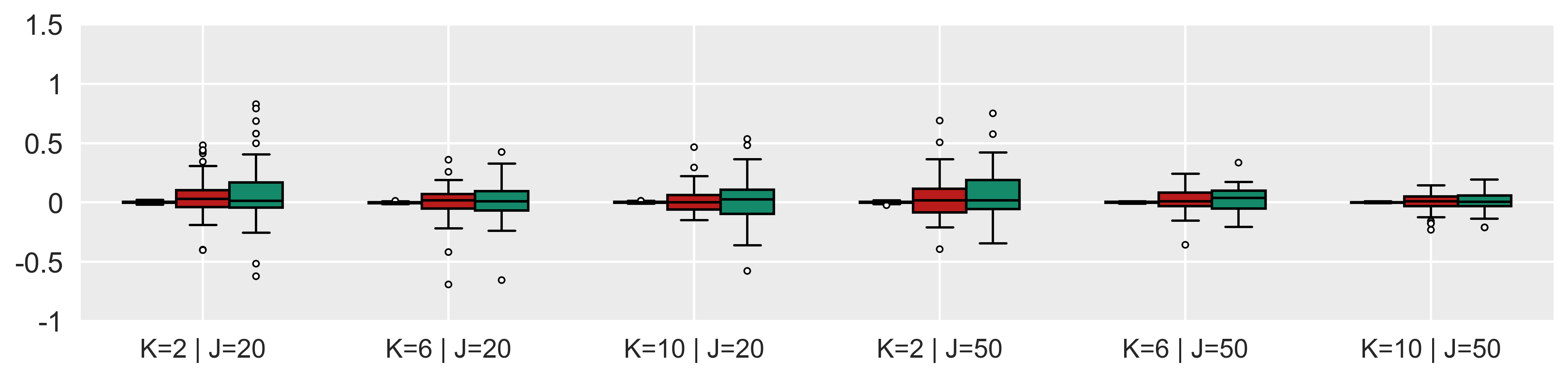}
        \caption{Bias for $b$}
        \label{fig:b_bias_mog_conf}
    \end{subfigure}

    \vspace{0.1em}

    \begin{subfigure}[b]{0.95\textwidth}
        \centering
        \includegraphics[width=\textwidth]{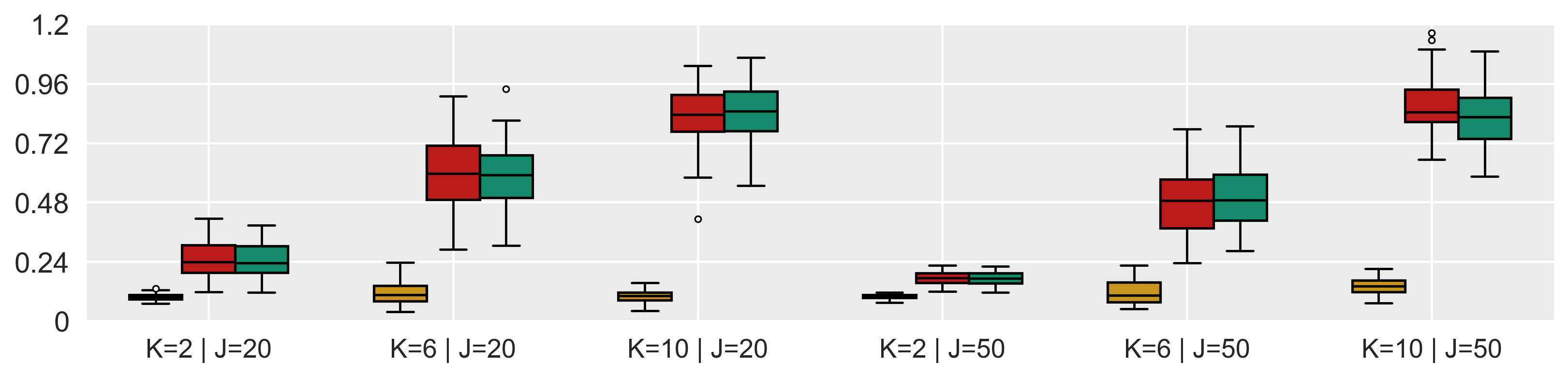}
        \caption{RMSE for $\alpha$}
        \label{fig:a_rmse_mog_conf}
    \end{subfigure}

    \vspace{0.1em}

    \begin{subfigure}[b]{0.95\textwidth}
        \centering
        \includegraphics[width=\textwidth]{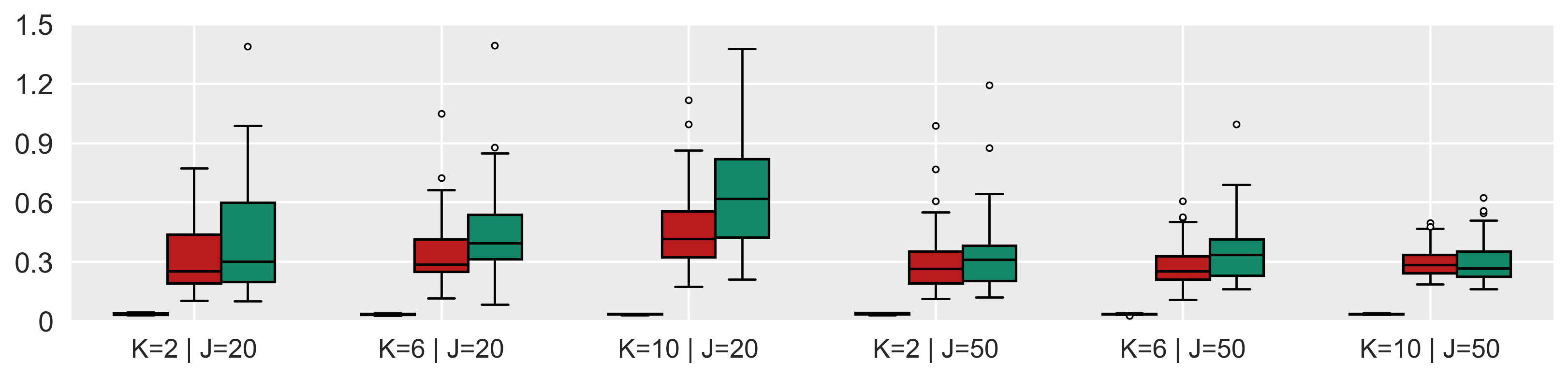}
        \caption{RMSE for $b$}
        \label{fig:b_rmse_mog_conf}
    \end{subfigure}

    \caption{Item parameter recovery for M2PL models under a mixture Gaussian latent distribution in the confirmatory setting, comparing MCEM, MHRM, and the proposed method.}
    \label{fig:mog_conf}
\end{figure}

\begin{figure}[!htbp]
    \centering
    
    \includegraphics[width=0.3\textwidth]{legend.png}
    \vspace{0.01em}
    \begin{subfigure}[b]{0.95\textwidth}
        \centering
        \includegraphics[width=\textwidth]{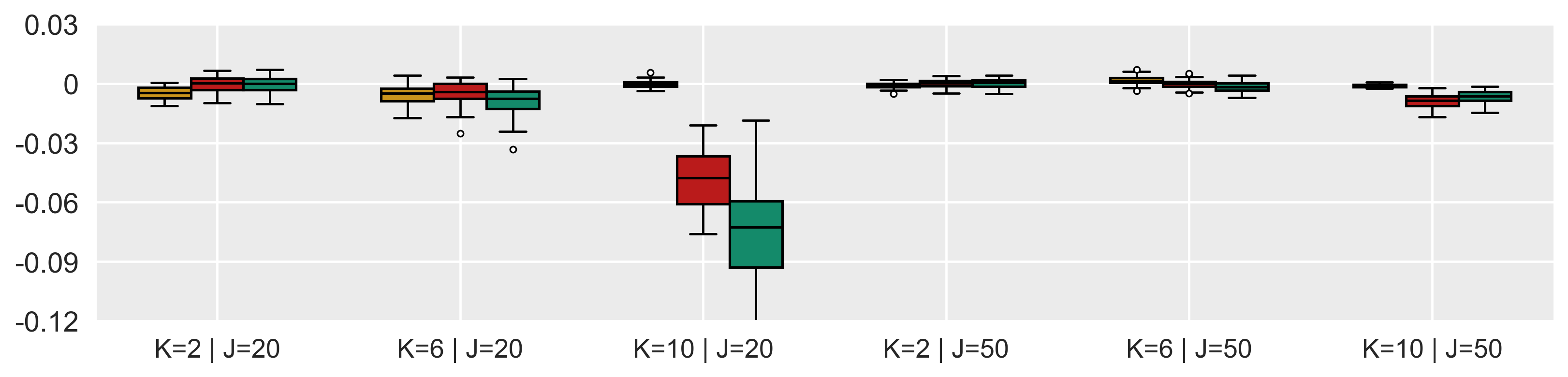}
        \caption{Bias for $\alpha$}
        \label{fig:a_bias_t_conf}
    \end{subfigure}

    \vspace{0.1em}

    \begin{subfigure}[b]{0.95\textwidth}
        \centering
        \includegraphics[width=\textwidth]{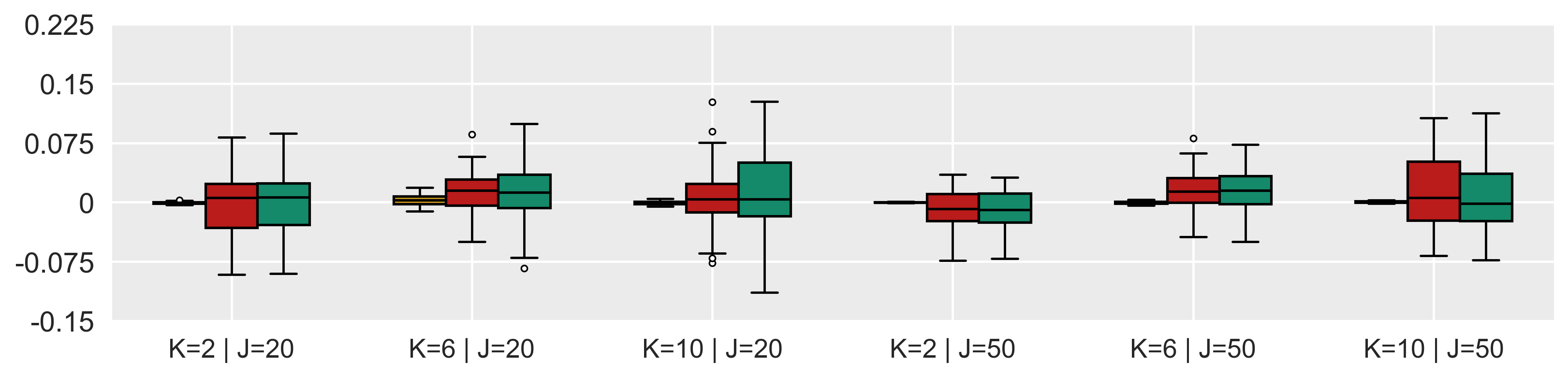}
        \caption{Bias for $b$}
        \label{fig:b_bias_t_conf}
    \end{subfigure}

    \vspace{0.1em}

    \begin{subfigure}[b]{0.95\textwidth}
        \centering
        \includegraphics[width=\textwidth]{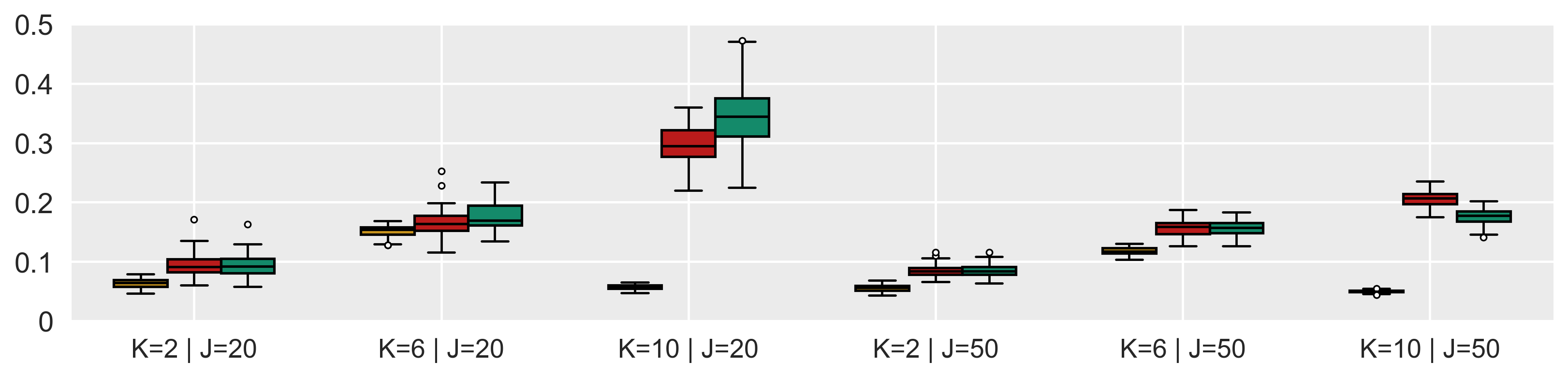}
        \caption{RMSE for $\alpha$}
        \label{fig:a_rmse_t_conf}
    \end{subfigure}

    \vspace{0.1em}

    \begin{subfigure}[b]{0.95\textwidth}
        \centering
        \includegraphics[width=\textwidth]{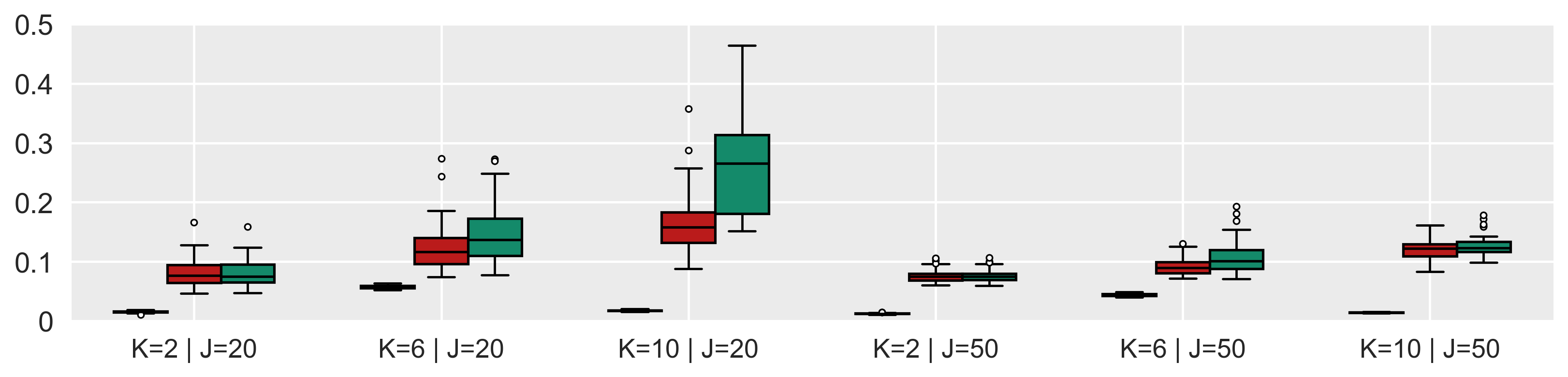}
        \caption{RMSE for $b$}
        \label{fig:b_rmse_t_conf}
    \end{subfigure}

    \caption{Item parameter recovery for M2PL models under $t$ distribution in the confirmatory setting, comparing MCEM, MHRM, and the proposed method.}
    \label{fig:t_conf}
\end{figure}}

{\spacingset{1.2}
\begin{figure}[H]
    \centering

    \includegraphics[width=0.25\linewidth]{legend.png}
    \vspace{0.5em}
        
\begin{subfigure}[t]{0.95\linewidth}
            \centering
            \includegraphics[width=\linewidth]{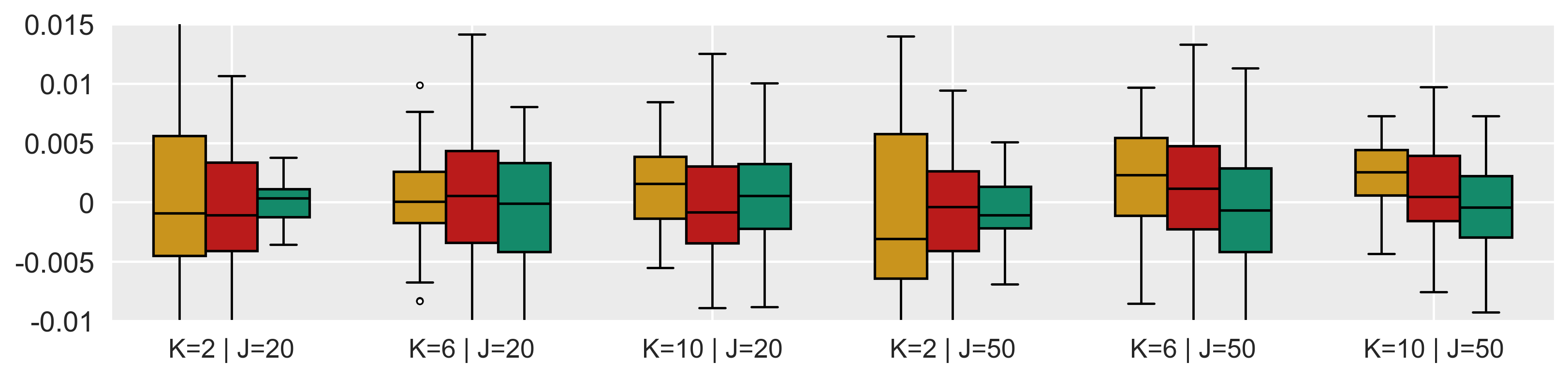}
            \caption{Gaussian}
            \label{fig:theta_bias_normal_conf_2}
        \end{subfigure}
        
        \vspace{0.5em}

        \begin{subfigure}[t]{0.95\linewidth}
            \centering
            \includegraphics[width=\linewidth]{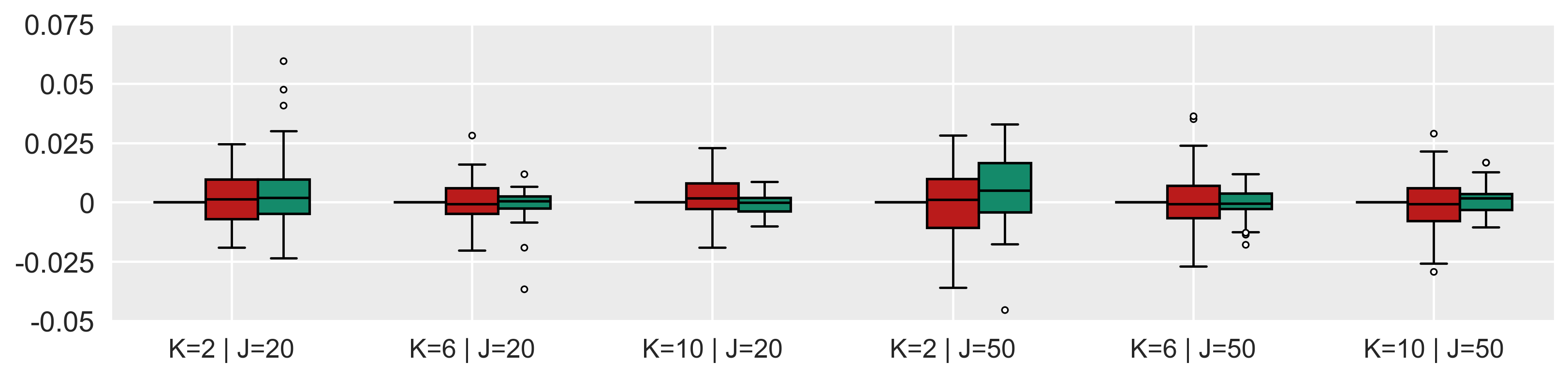}
            \caption{Mixture of Gaussian}
            \label{fig:theta_bias_mog_conf}
        \end{subfigure}

        \vspace{0.5em}

        \begin{subfigure}[t]{0.95\linewidth}
            \centering
            \includegraphics[width=\linewidth]{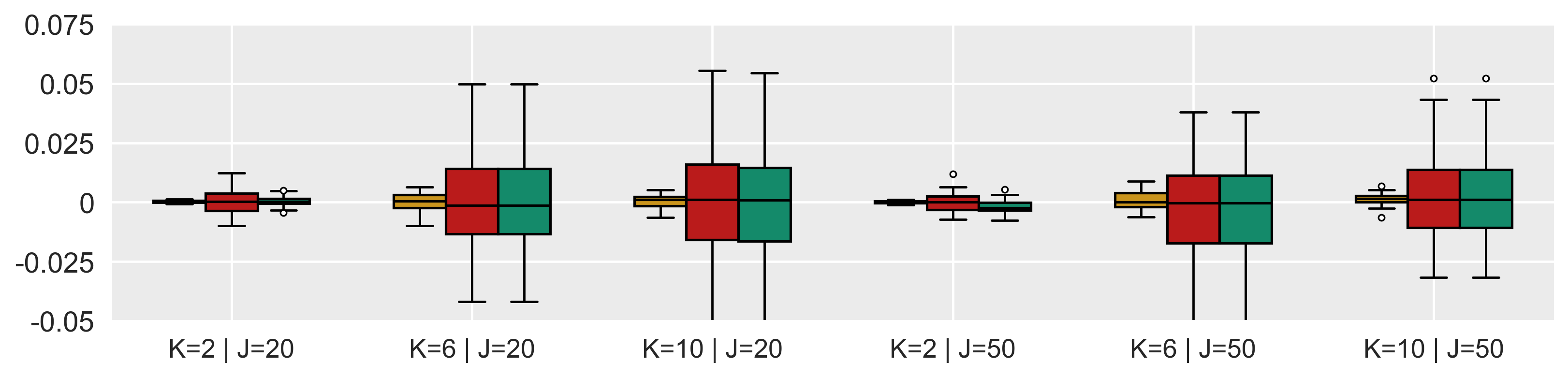}
            \caption{Student-\(t\)}
            \label{fig:theta_bias_t_conf}
        \end{subfigure}

    \caption{Bias for latent trait recovery in the confirmatory setting. Rows correspond to Gaussian, Mixture of Gaussian, and Student-\(t\) latent distributions.}
    \label{fig:appendix_theta_conf_bias}
\end{figure}}

The results are similar to those in the exploratory setting. When the latent traits follow a Gaussian distribution, the proposed method performs comparably to MCEM and MHRM in terms of item parameter recovery, although it is slightly less accurate in some settings, such as when $K=2$; see Figure~\ref{fig:normal_conf}. When the latent distribution is non-Gaussian, the proposed method shows clear advantages over MCEM and MHRM in item parameter estimation, as illustrated in Figures~\ref{fig:mog_conf} and~\ref{fig:t_conf}.
The latent trait recovery results exhibit similar patterns. Figures~\ref{fig:appendix_theta_conf_bias} and~\ref{fig:appendix_theta_conf_rmse} show that the proposed method achieves better performance in both bias and RMSE when the latent distribution is non-Gaussian. Under the Gaussian setting, its performance is slightly worse in some cases but remains comparable to the benchmark methods.

{\spacingset{1.2}
\begin{figure}[H]
    \centering

    \includegraphics[width=0.25\linewidth]{legend.png}
    \vspace{0.5em}
        \begin{subfigure}[t]{0.95\linewidth}
            \centering
            \includegraphics[width=\linewidth]{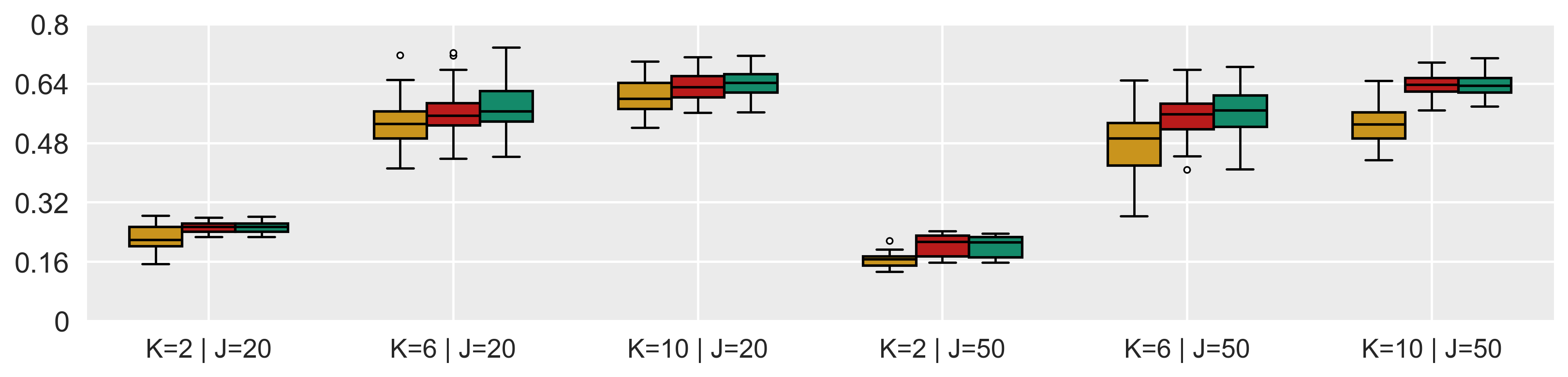}
            \caption{Mixture of Gaussian}
            \label{fig:theta_rmse_mog_conf}
        \end{subfigure}

        \vspace{0.5em}

        \begin{subfigure}[t]{0.95\linewidth}
            \centering
            \includegraphics[width=\linewidth]{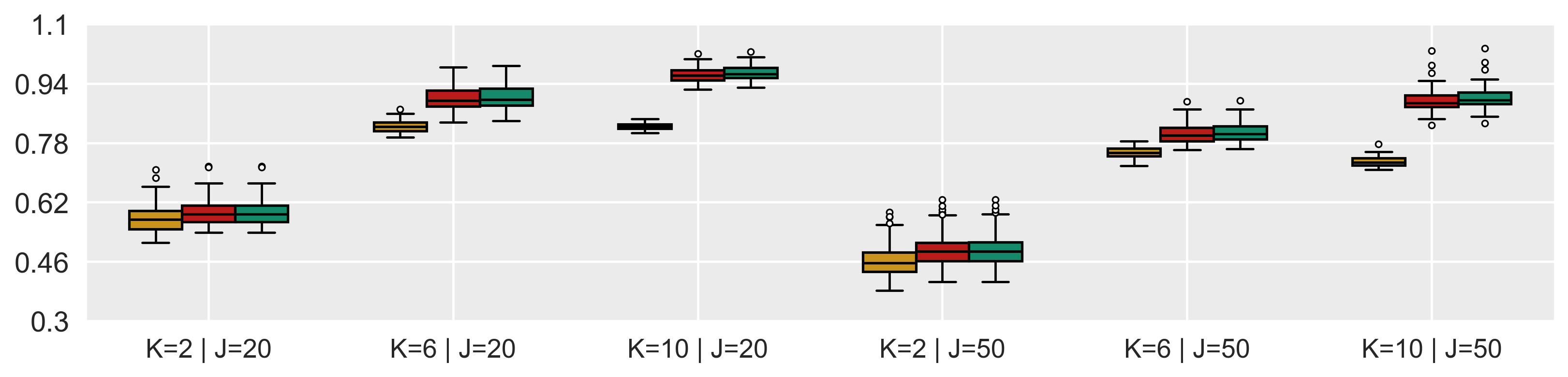}
            \caption{Student-\(t\)}
            \label{fig:theta_rmse_t_conf}
        \end{subfigure}
\begin{subfigure}[t]{0.95\linewidth}
            \centering
            \includegraphics[width=\linewidth]{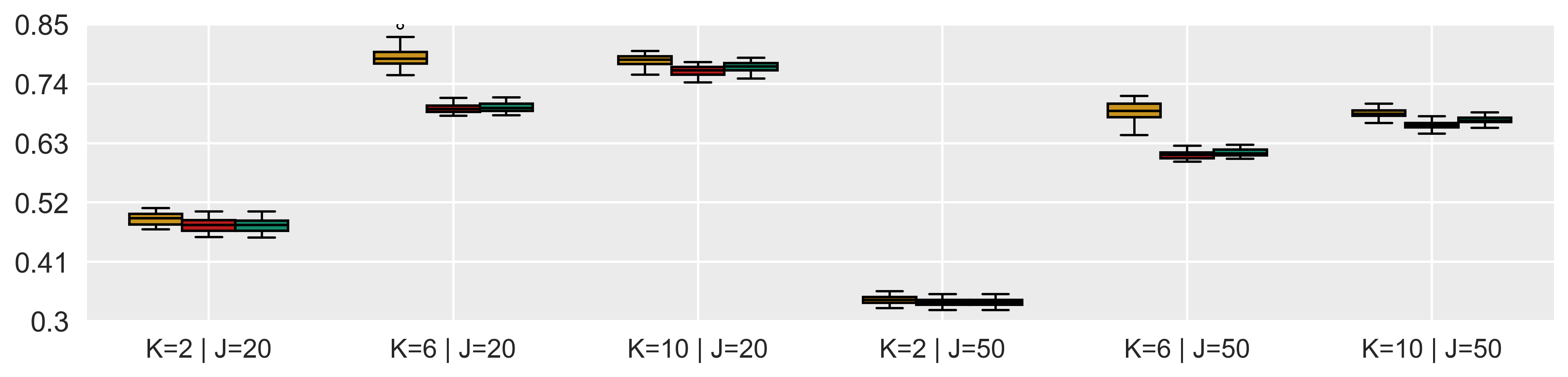}
            \caption{Gaussian}
            \label{fig:theta_rmse_normal_conf}
        \end{subfigure}
        \vspace{0.5em}

    \caption{RMSE for latent trait recovery in the confirmatory setting. Rows correspond to Gaussian, Mixture of Gaussian, and Student-\(t\) latent distributions.}
    \label{fig:appendix_theta_conf_rmse}
\end{figure}}

Overall, these results suggest that the proposed method provides a flexible framework for MIRT estimation when the latent distribution is unknown or varies across applications. It remains competitive when the Gaussian assumption is approximately correct and provides substantial gains when the latent distribution departs from normality. This makes the proposed method particularly useful in practical settings where the exact true latent distribution is rarely known in advance, and especially when prior knowledge suggests potential non-Gaussian features.

\section{Real Data Analysis}\label{sec_data}

In this section, we analyze the Big Five personality dataset, which is publicly available at
\href{https://openpsychometrics.org/tests/IPIP-BFFM/}{https://openpsychometrics.org/tests/IPIP-BFFM/}.
The Big Five personality factor model is one of the most widely used frameworks for describing personality in psychology~\citep{goldberg1992development}. The dataset was collected through an online Big Five personality assessment conducted in 2012 and contains 19,718 observations. In our analysis, we restrict attention to participants from the United States, yielding a sample of 8,753 individuals.

The questionnaire consists of 50 items designed to measure five personality traits: Extraversion, Neuroticism, Agreeableness, Conscientiousness, and Openness. Each item is rated on a five-point Likert scale, where 1 indicates \textit{Disagree}, 3 indicates \textit{Neutral}, and 5 indicates \textit{Agree}; a response of 0 indicates a missing value. To analyze the data under the binary response setting of the multidimensional two-parameter logistic (M2PL) model, we dichotomize the responses by mapping $\{1,2,3\}$ to 1 and $\{4,5\}$ to 0. We apply the proposed method with latent dimension $K = 5$ to estimate the item parameters, the latent distributions for the five dimensions, and the posterior approximation.

The transformation for the conditional density $g_{\phi}(\cdot)$ is implemented using a similar neural network architecture as in the simulation studies in Section~\ref{sec_simu}, with the output dimension set to \(K=5\). To assess whether the Gaussian latent distribution assumption is adequate, we apply the summed-score likelihood-based test~\citep{li2018summed}. The test result rejects the null hypothesis of normality with a p-value of $1.91\times 10^{-18}$, indicating that the latent trait distribution deviates from Gaussianity. Therefore, we use a relatively expressive neural spline flow for $h_{\eta}(\cdot)$ with 10 layers, hidden dimension 256, 24 spline bins, and tail bound 12.0 to capture potential non-Gaussian features in the latent trait distribution.

{\spacingset{1.2}\begin{figure}[!htbp]
    \centering
    \makebox[\textwidth][c]{%
        \includegraphics[
            height=1\textwidth,
            angle=-90
        ]{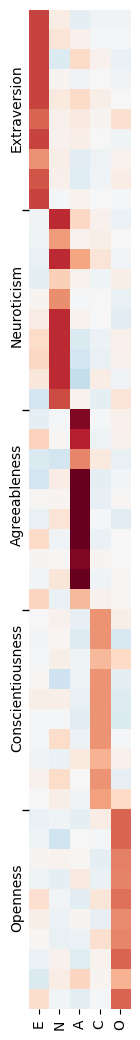}
    }
    \caption{Heatmap of the estimated transposed discrimination matrix
$A^\top=(\alpha_1,\ldots,\alpha_{50})$ for the U.S. Big Five data. Rows correspond to the five latent personality factors: Extraversion (E), Neuroticism (N), Agreeableness (A), Conscientiousness (C), and Openness (O). Columns correspond to items, grouped by their intended Big Five domains.}
    \label{heatmap}
\end{figure}}

Figure~\ref{heatmap} presents a heatmap of the estimated discrimination parameters. The columns correspond to questionnaire items grouped by their intended Big Five domains: Extraversion, Neuroticism, Agreeableness, Conscientiousness, and Openness. The rows correspond to the five estimated latent personality traits, labeled by their initials: E, N, A, C, and O. The heatmap exhibits a clear block structure in which item responses within each domain are largely dependent on one primary latent trait, while their dependence on the remaining traits is comparatively small. This pattern indicates that the estimated latent traits align well with the intended Big Five domains, while still allowing for modest cross-domain associations.


{\spacingset{1.2}
\begin{figure}[!htbp]
    \centering
    \captionsetup[subfigure]{font=footnotesize}
    \captionsetup{font=small, skip=3pt}

    \includegraphics[width=0.35\textwidth]{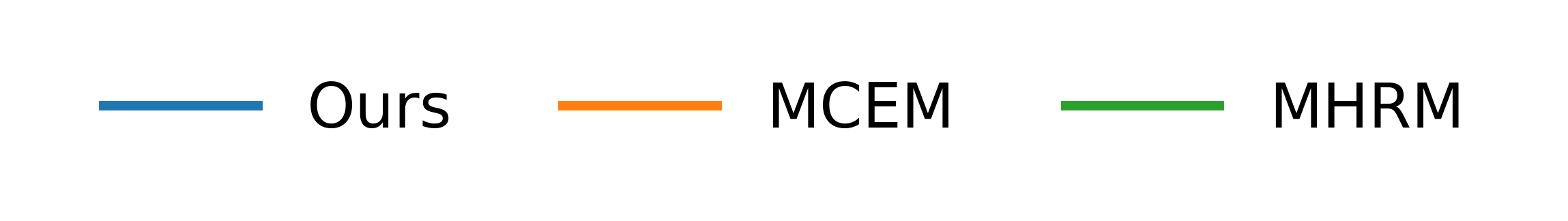}
    \vspace{-0.5em}

    \begin{subfigure}[b]{0.3\textwidth}
        \centering
        \includegraphics[width=\textwidth]{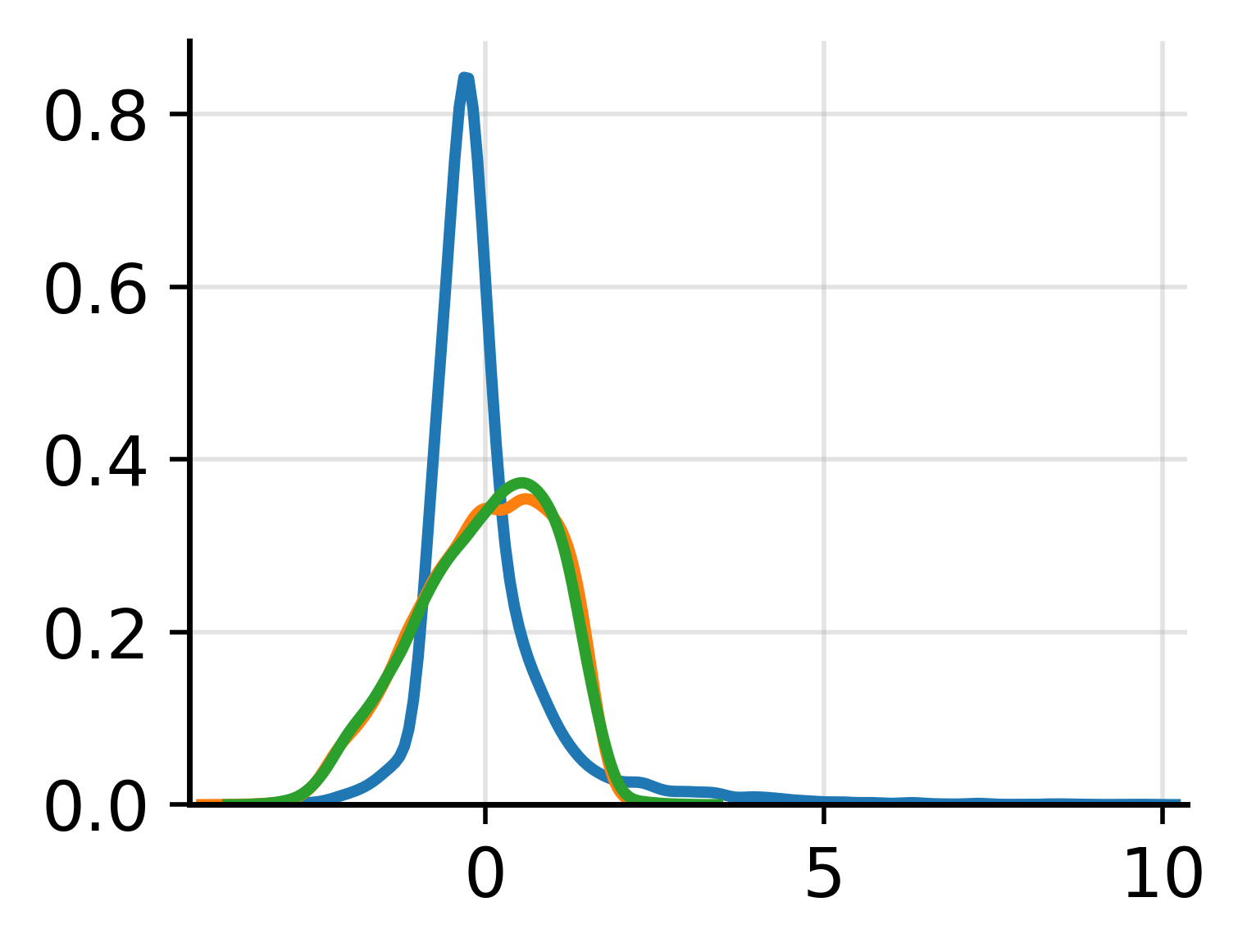}
        \caption{$\theta_1$ (Agreeableness)}
        \label{fig:hist_z1}
    \end{subfigure}
    \hfill
    \begin{subfigure}[b]{0.3\textwidth}
        \centering
        \includegraphics[width=\textwidth]{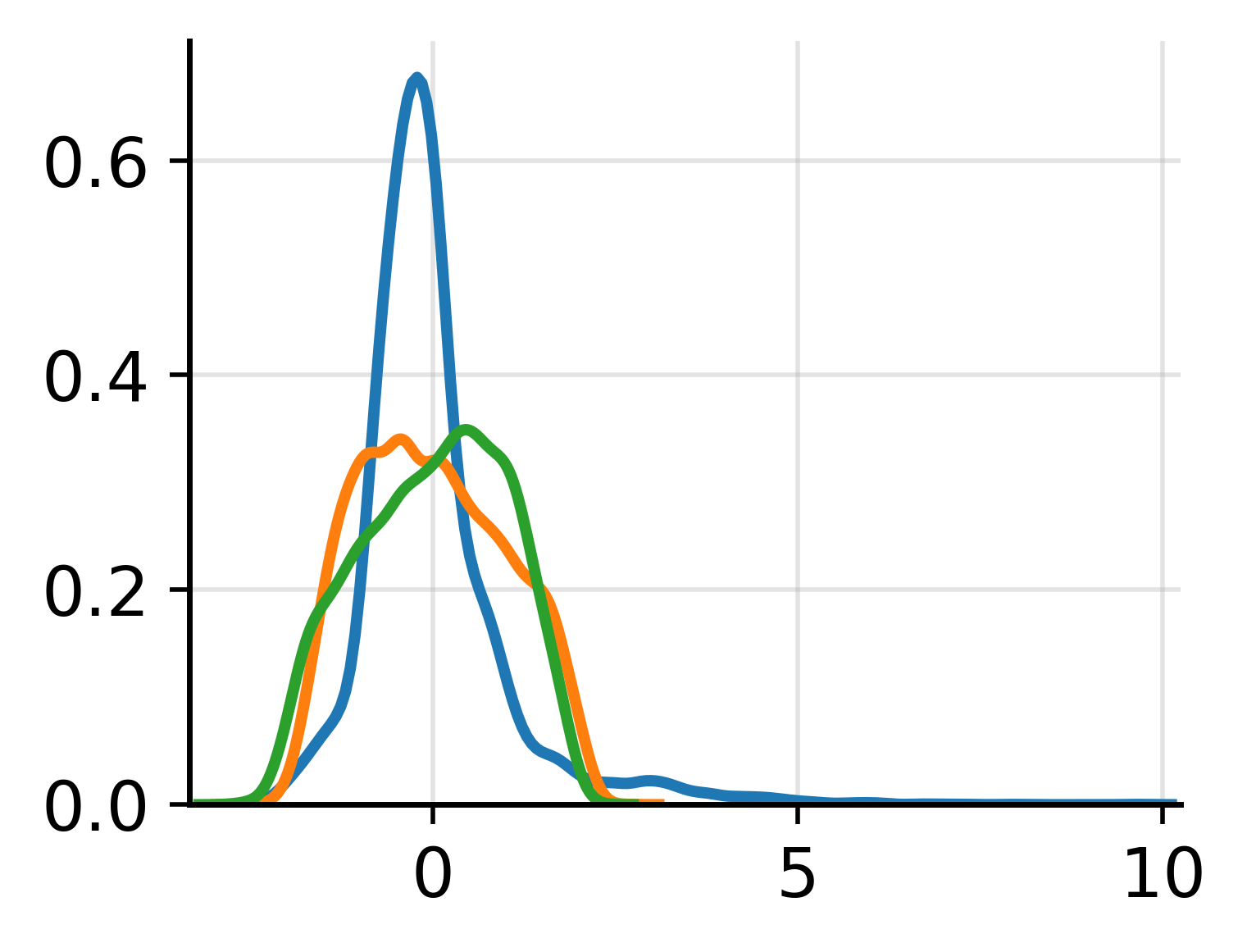}
        \caption{$\theta_2$ (Openness)}
        \label{fig:hist_z2}
    \end{subfigure}
    \hfill
    \begin{subfigure}[b]{0.3\textwidth}
        \centering
        \includegraphics[width=\textwidth]{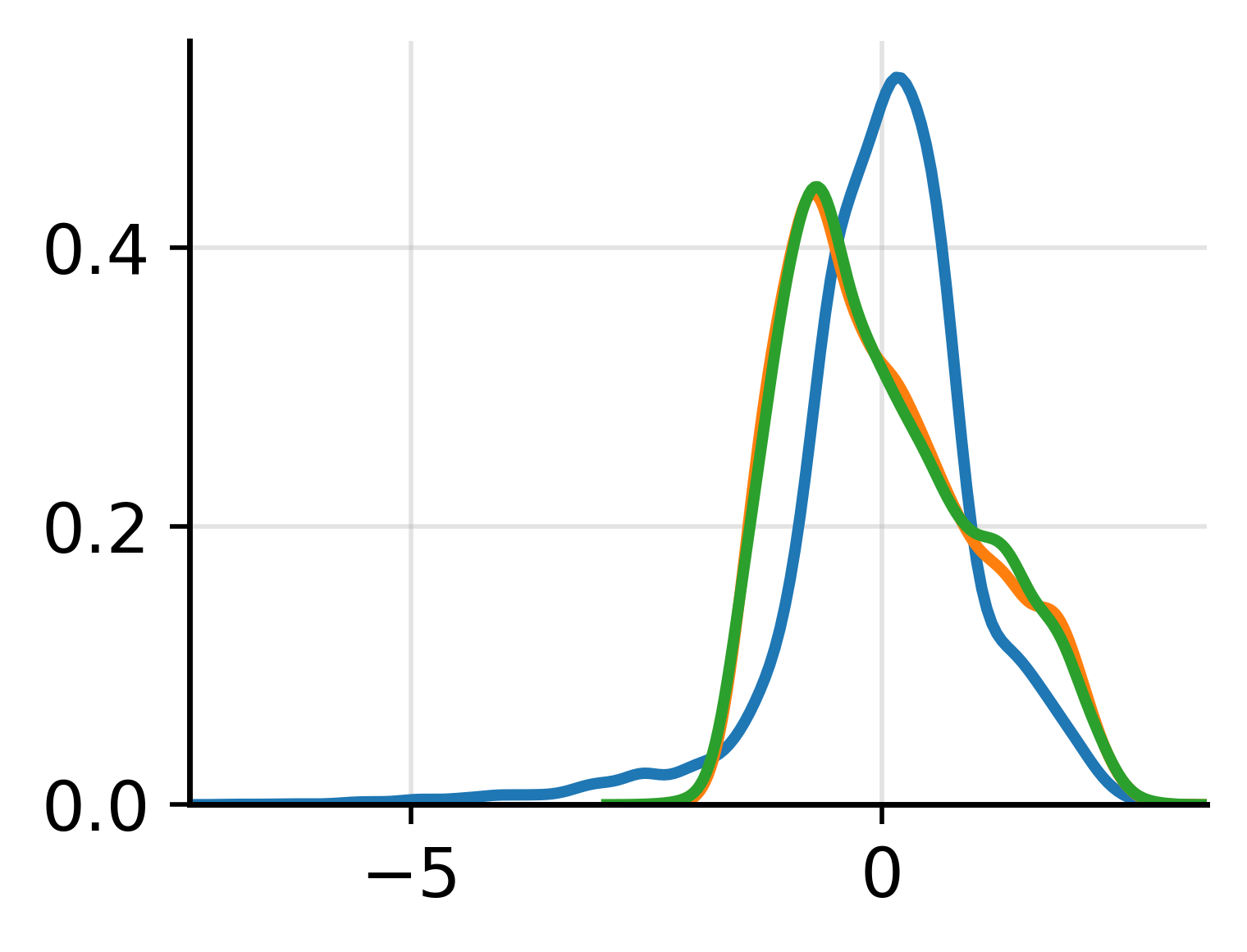}
        \caption{$\theta_3$ (Conscientiousness)}
        \label{fig:hist_z3}
    \end{subfigure}
    \hfill
    \begin{subfigure}[b]{0.3\textwidth}
        \centering
        \includegraphics[width=\textwidth]{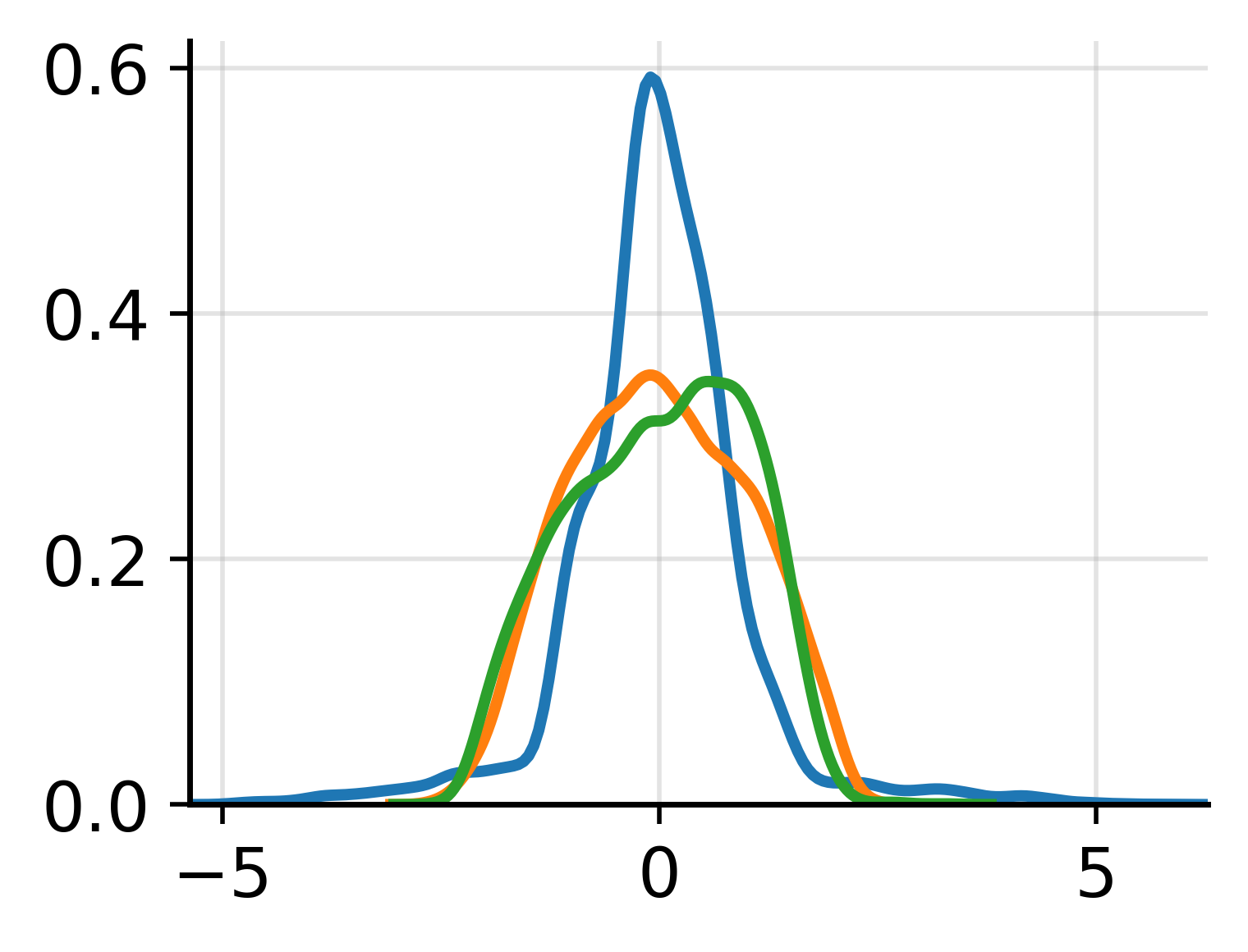}
        \caption{$\theta_4$ (Extraversion)}
        \label{fig:hist_z4}
    \end{subfigure}
    \begin{subfigure}[b]{0.3\textwidth}
        \centering
        \includegraphics[width=\textwidth]{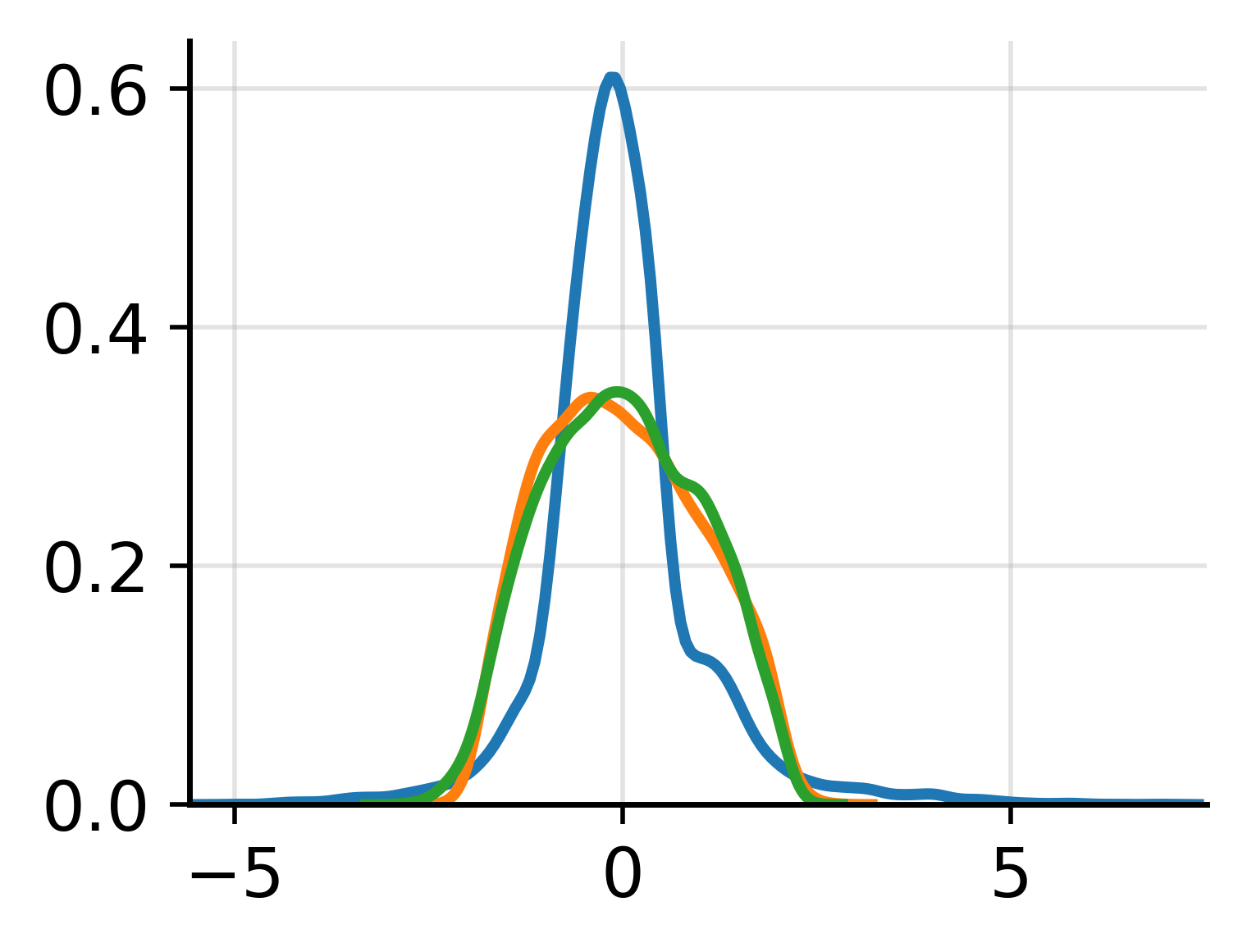}
        \caption{$\theta_5$ (Neuroticism)}
        \label{fig:hist_z5}
    \end{subfigure}

    \vspace{-0.4em}
    \caption{Estimated distributions for the five personality factors in the Big Five data.}
    \label{fig:hist_latent}
\end{figure}

\begin{figure}[!htbp]
    \centering
    \captionsetup{font=small, skip=3pt}
    \captionsetup[subfigure]{font=footnotesize}

    \includegraphics[width=0.65\textwidth]{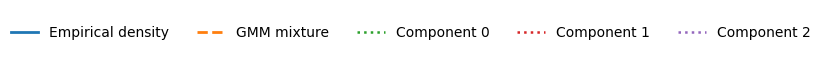}
    \vspace{-0.6em}

    \begin{subfigure}[b]{0.3\textwidth}
        \centering
        \includegraphics[width=\textwidth]{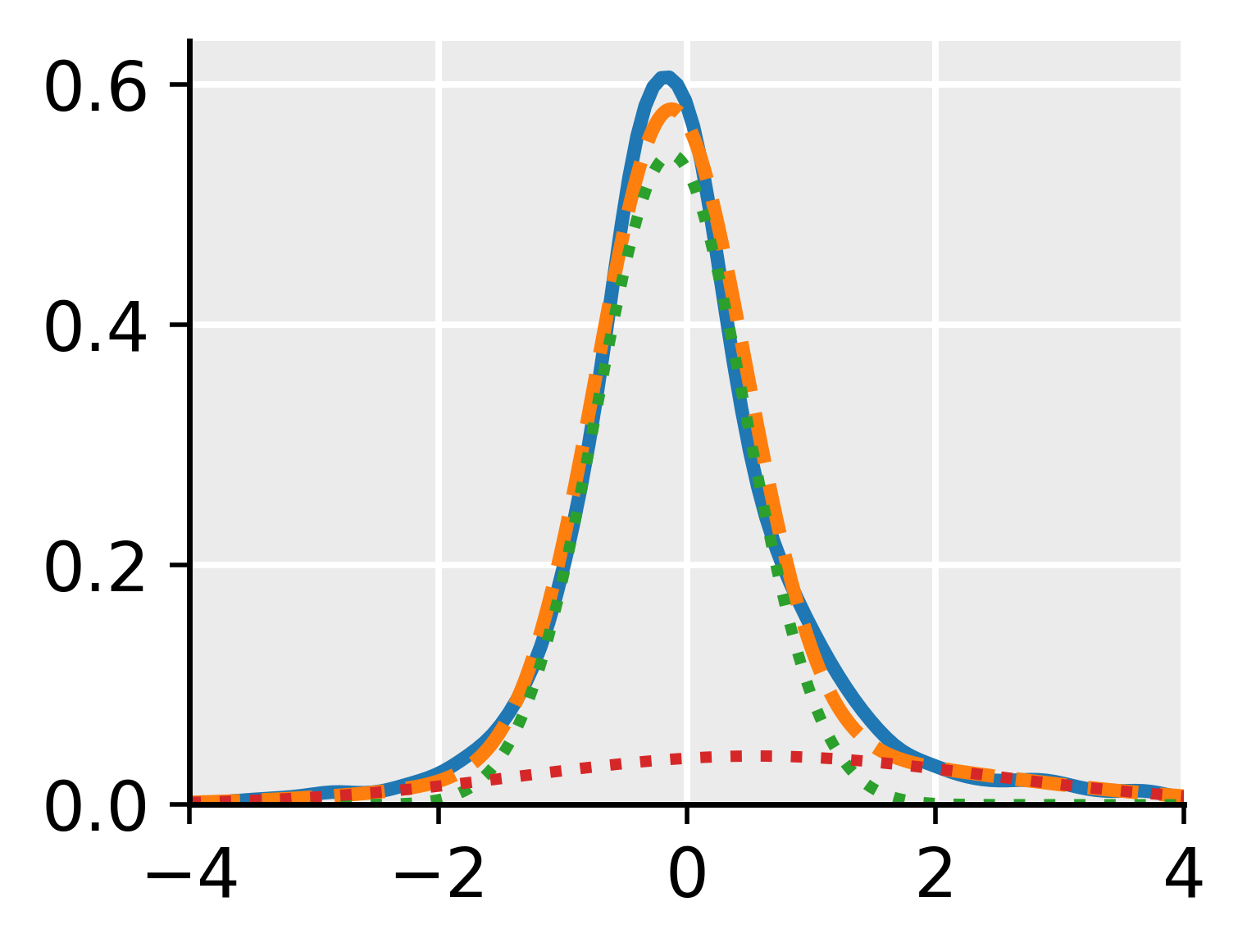}
        \caption{$\theta_1$ (Agreeableness)}
        \label{fig:itemA_z1}
    \end{subfigure}
    \hfill
    \begin{subfigure}[b]{0.3\textwidth}
        \centering
        \includegraphics[width=\textwidth]{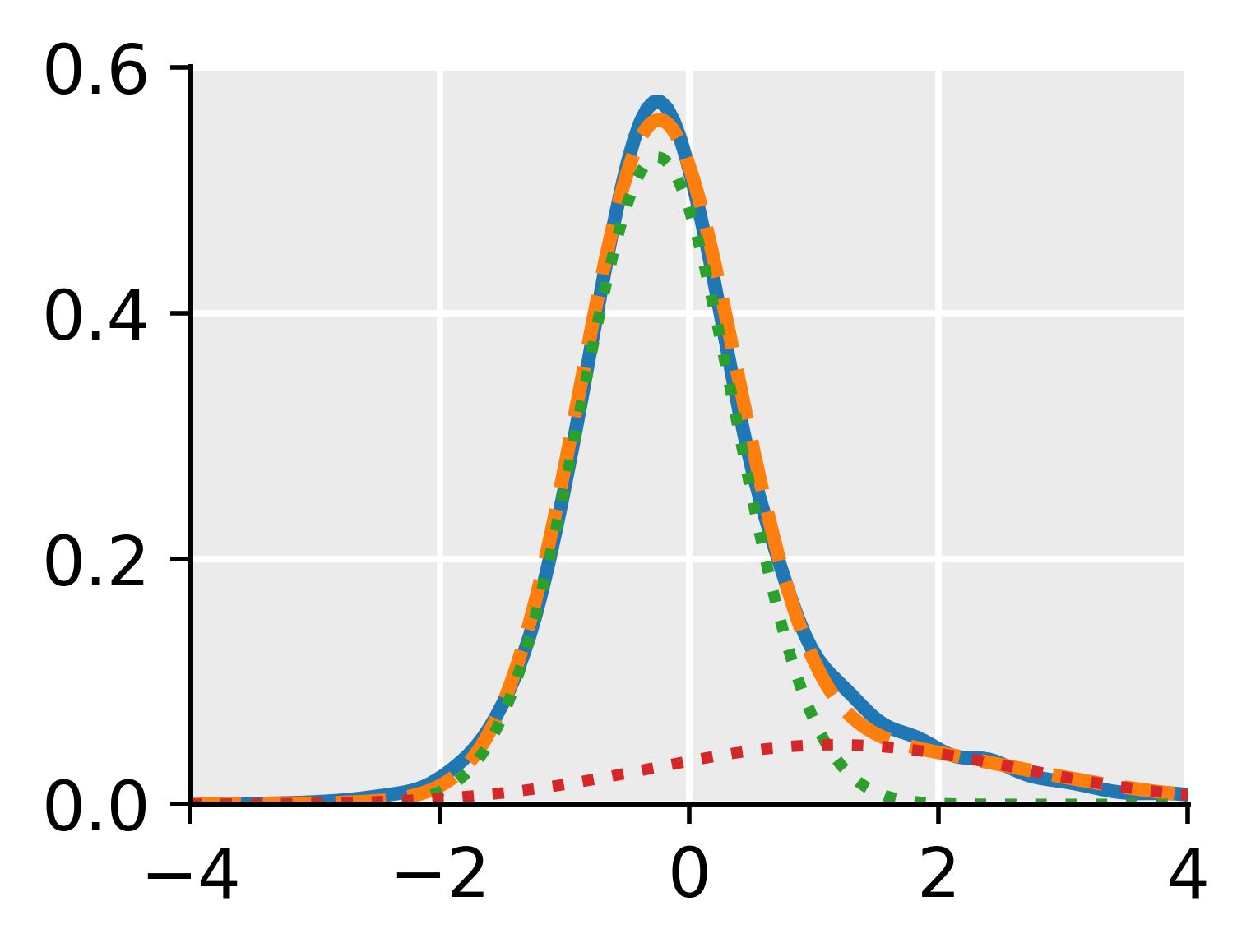}
        \caption{$\theta_2$ (Openness)}
        \label{fig:itemO_z2}
    \end{subfigure}
    \hfill
    \begin{subfigure}[b]{0.3\textwidth}
        \centering
        \includegraphics[width=\textwidth]{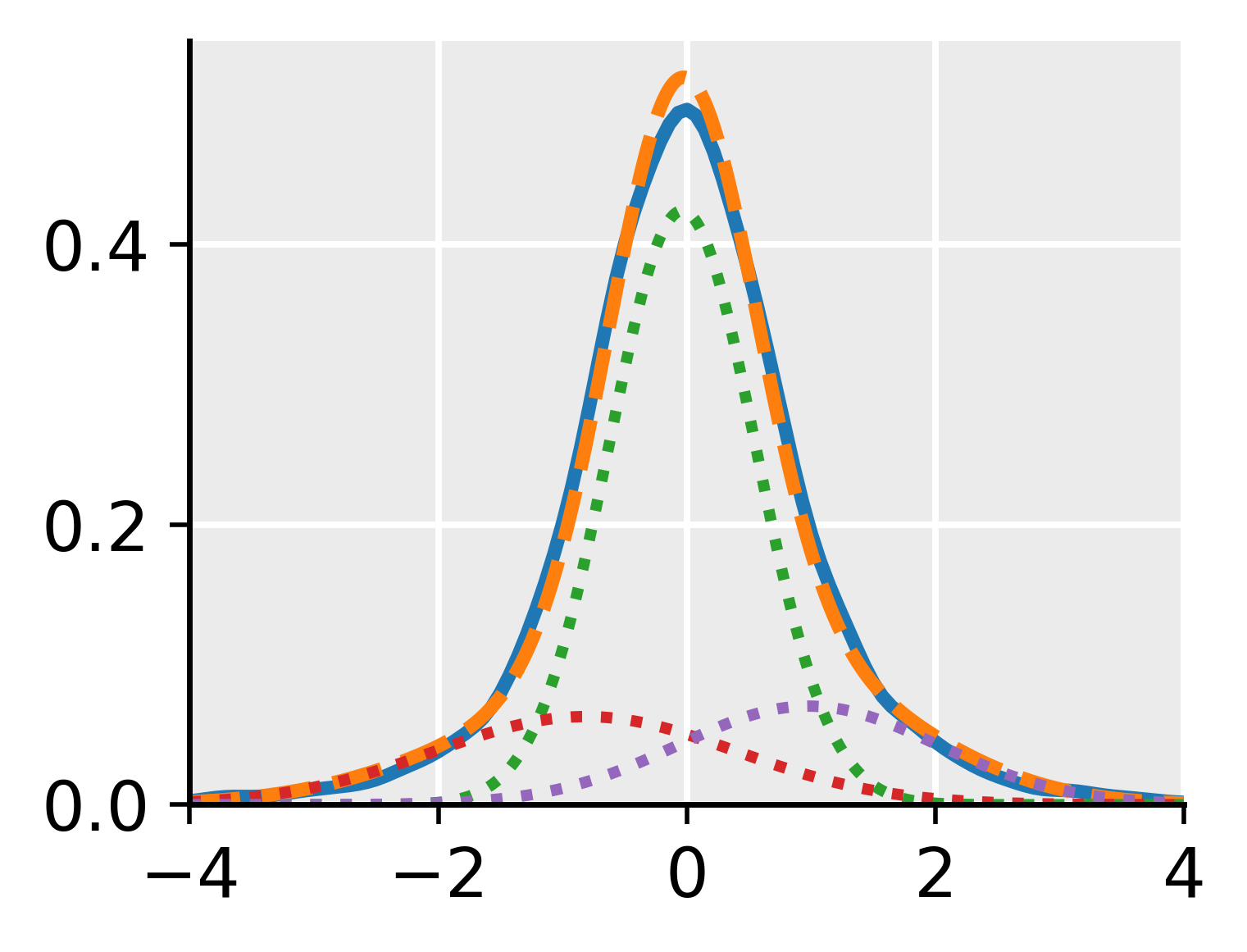}
        \caption{$\theta_3$ (Conscientiousness)}
        \label{fig:itemC_z3}
    \end{subfigure}
    \hfill
    \begin{subfigure}[b]{0.3\textwidth}
        \centering
        \includegraphics[width=\textwidth]{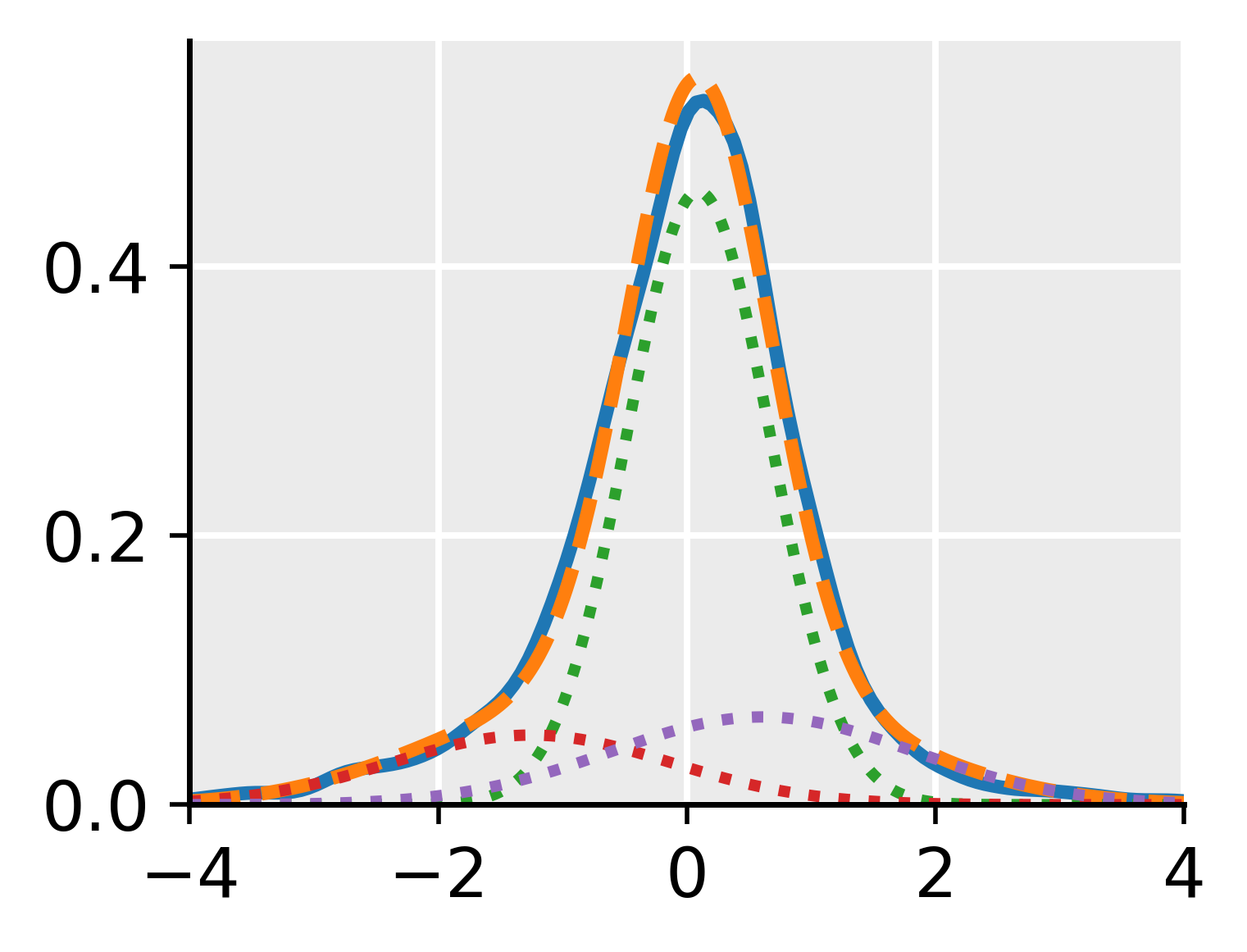}
        \caption{$\theta_4$ (Extraversion)}
        \label{fig:itemE_z4}
    \end{subfigure}
    \begin{subfigure}[b]{0.3\textwidth}
        \centering
        \includegraphics[width=\textwidth]{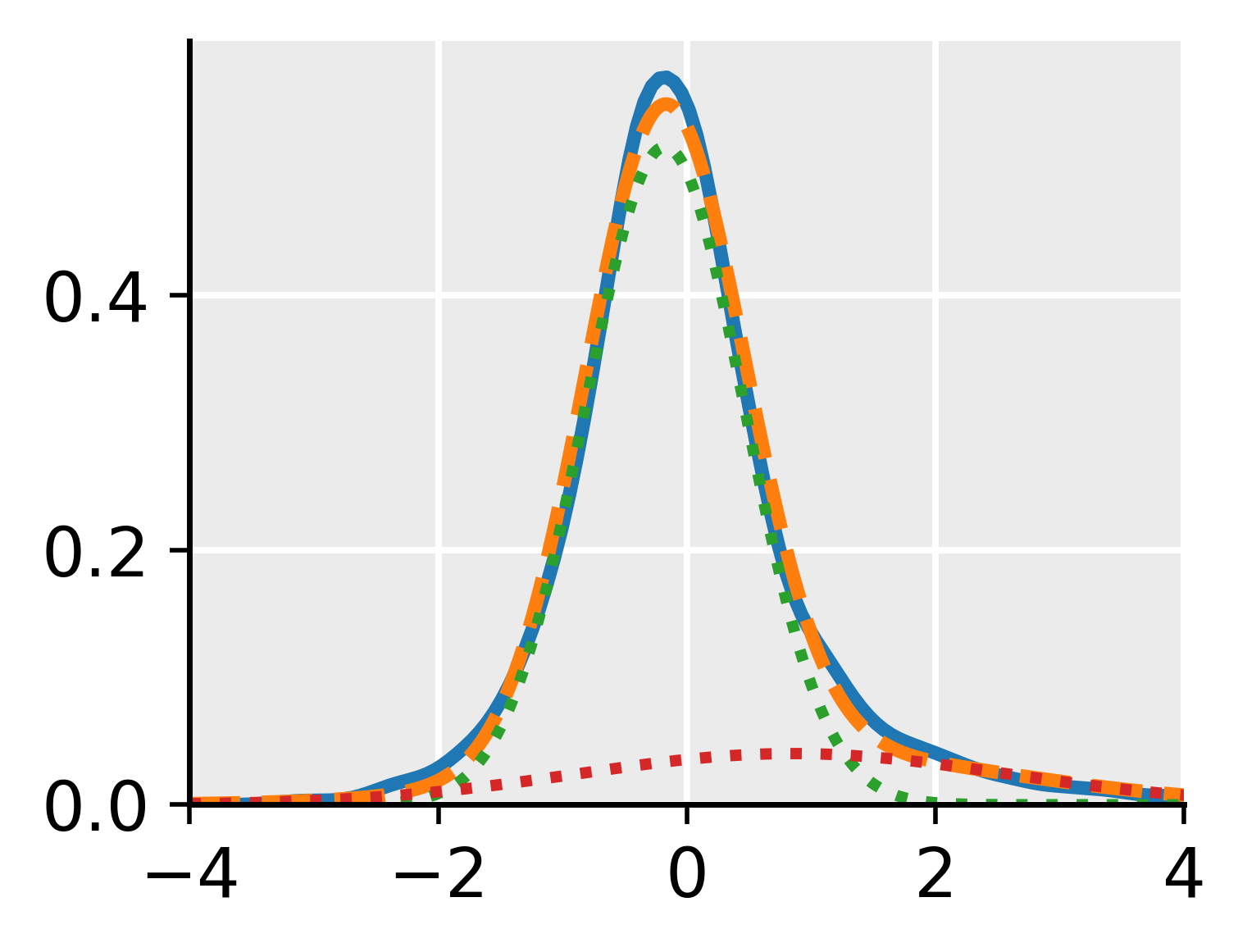}
        \caption{$\theta_5$ (Neuroticism)}
        \label{fig:itemN_z5}
    \end{subfigure}
    \vspace{-0.4em}
    \caption{Gaussian mixture approximations to the estimated latent trait distributions for the Big5 data. Each panel shows the empirical density, fitted mixture density, and component densities; two components are used for $\theta_1$, $\theta_2$, and $\theta_5$, and three components are used for $\theta_3$ and $\theta_4$.}
    \label{fig:item_cluster_examples}
\end{figure}}

Figure~\ref{fig:hist_latent} displays the estimated latent trait distributions across the five dimensions for the proposed method, MCEM, and MHRM. While MCEM and MHRM produce distributions that are approximately Gaussian in shape, the latent traits estimated by the proposed method exhibit clear departures from normality. Consistent with the likelihood-based test, which rejects the null hypothesis of a Gaussian latent density, these results suggest that the latent trait distribution in this dataset is not well captured by a multivariate normal model. Additional QQ plots illustrating these deviations are provided in Appendix~\ref{supp_real_data}.

Figure~\ref{fig:item_cluster_examples} presents Gaussian mixture approximations to the estimated latent factor distributions for the five latent dimensions. 
For dimensions with three mixture components ($\theta_3$ and $\theta_4$), the estimated distributions typically contain one central component near zero and two additional components on the negative and positive sides. For example, $\theta_3$, which represents Conscientiousness, separates individuals into groups with low, neutral, and high levels of the trait. A similar pattern is observed for $\theta_4$, corresponding to Extraversion, where the groups may be interpreted as introverted, neutral, and extraverted individuals.
For dimensions with two mixture components  ($\theta_1$, $\theta_2$, and $\theta_5$), the central component dominates, while individuals with more extreme trait values form a smaller secondary component. These patterns suggest again that the population is not well characterized by a single symmetric Gaussian distribution, but instead exhibits heterogeneous subgroups with varying trait intensities. This finding highlights the need for estimation methods that allow flexible latent distributions, as empirical latent trait distributions can be complex and may vary substantially across dimensions.

Table~\ref{tab:item_factor_cluster_summary} further summarizes the response patterns within the latent clusters identified from the Gaussian mixture approximations. The results show that these clusters are not merely distributional artifacts, but correspond to clearly different response profiles on the original Likert scale. For example, for the Agreeableness trait $\theta_1$, cluster 0 has a higher mean response than cluster 1 ($4.223$ versus $3.870$), with a substantially larger proportion of high responses ($0.825$ versus $0.689$). A similar pattern is observed for the Openness trait $\theta_2$, where the smaller cluster is associated with stronger agreement responses, with nearly 89\% of responses falling in categories 4--5. For the Conscientiousness and Extraversion traits ($\theta_3$ and $\theta4$), the mixture components reveal more pronounced heterogeneity. In particular, the three clusters for $\theta_3$ correspond to relatively high, moderate, and low response groups, with mean responses $4.015$, $3.405$, and $3.070$, respectively. For $\theta_4$, the separation is even stronger: cluster 0 is dominated by high responses, with a mean response of $4.259$ and 85\% of responses in categories 4--5, whereas cluster 1 is dominated by low responses, with a mean response of $2.303$ and 61\%
of responses in categories 1--2. Cluster 2 lies between these two groups, suggesting a neutral or intermediate subgroup. The Neuroticism trait $\theta_5$ also shows a meaningful two-cluster structure. Compared with the dominant cluster, the smaller cluster has a lower mean response ($2.472$ versus $3.099$) and a much larger proportion of low responses (59\% versus 35\%). 
Taken together, these cluster-level summaries show that the non-Gaussian features of the estimated latent distributions are reflected in the observed response patterns. The mixture components correspond to respondents with systematically different response tendencies on the relevant item groups, rather than to purely numerical artifacts of the density approximation. This indicates the substantive practical relevance of the flexible latent distribution learned by the proposed method.

\begin{table}[!htbp]
    \centering
    \captionsetup{font={small,stretch=1.2}, skip=3pt}
    \caption{ Cluster-based response summaries across the five item domains. 
    $\theta_1$--$\theta_5$ denote the personality traits Agreeableness (A), Openness (O), Conscientiousness (C), Extraversion (E), and Neuroticism (N), respectively. The columns labeled 1--5 report the percentages of responses in each Likert category within each cluster. The Low and High columns report the aggregated percentages of low responses, defined as categories 1--2, and high responses, defined as categories 4--5, respectively.}
    \label{tab:item_factor_cluster_summary}

    \resizebox{0.98\textwidth}{!}{
    \begin{tabular}{llcccccccc}
    \hline
    Factor & Cluster & Mean Response & 1 (\%) & 2 (\%) & 3 (\%) & 4 (\%) & 5 (\%) & Low (1--2, \%) & High (4--5, \%)\\
    \hline
    \multirow{2}{*}{$\theta_1$}
    & 0 & 4.223 & 3.2 & 5.7 & 8.6 & 30.5 & 52.0 & 8.9 & 82.5  \\
    & 1 & 3.870 & 4.0 & 9.1 & 17.9 & 33.7 & 35.2 & 13.1 & 68.9  \\
    \hline
    \multirow{2}{*}{$\theta_2$}
    & 0 & 3.887 & 3.5 & 7.9 & 20.5 & 33.0 & 35.3 & 11.3 & 68.2  \\
    & 1 & 4.406 & 1.6 & 2.3 & 6.7 & 32.8 & 56.6 & 3.9 & 89.4  \\
    \hline
    \multirow{3}{*}{$\theta_3$}
    & 0 & 3.405 & 8.7 & 15.3 & 24.5 & 30.0 & 21.5 & 23.9 & 51.5  \\
    & 1 & 4.015 & 4.2 & 7.6 & 11.9 & 35.1 & 41.2 & 11.8 & 76.3  \\
    & 2 & 3.070 & 12.6 & 21.6 & 26.8 & 24.3 & 14.7 & 34.2 & 39.0  \\
    \hline
    \multirow{3}{*}{$\theta_4$}
    & 0 & 4.259 & 3.9 & 4.1 & 6.7 & 32.7 & 52.6 & 8.0 & 85.2  \\
    & 1 & 2.303 & 32.3 & 28.7 & 22.1 & 10.2 & 6.7 & 61.0 & 16.9  \\
    & 2 & 3.010 & 16.6 & 20.6 & 23.7 & 23.5 & 15.7 & 37.2 & 39.1  \\
    \hline
    \multirow{2}{*}{$\theta_5$}
    & 0 & 3.099 & 13.5 & 21.9 & 22.9 & 24.8 & 17.0 & 35.4 & 41.8  \\
    & 1 & 2.472 & 28.8 & 30.4 & 17.1 & 12.2 & 11.5 & 59.2 & 23.7  \\
    \hline
    \end{tabular}
    }
\end{table}
\section{Concluding Remarks}\label{sec_conclude}

In this paper, we propose a flexible estimation approach for multidimensional item response theory (MIRT) models. Unlike conventional MIRT methods that rely on Gaussian assumptions for the latent traits, our method uses a flow-based representation of the latent distribution, allowing it to capture complex structures such as skewness, heavy tails, and multimodality. For estimation, we introduce a flow-based approximation to the posterior distribution as a function of both the observed responses and Gaussian noise. This framework enables efficient estimation of the item parameters, the latent distribution, and the posterior approximation. Extensive simulation studies and an application to real data demonstrate the effectiveness and practical applicability of the proposed method.

Our work opens several directions for future research. First, the proposed approach models the latent distribution through a transformation of Gaussian noise. In practice, covariate information may be incorporated to explain variation in the latent traits and to improve interpretability. For example, one may allow the flow transformation to depend on observed individual demographic features, leading to a conditional latent distribution that captures heterogeneity across subpopulations. Second, the specification of the prior and posterior flows can influence model fit and computational efficiency. Intuitively, more complex latent distributions may require richer parameterizations of $\eta$ and $\phi$ to adequately capture their distributional features. It remains an important open problem to develop data-driven approaches for selecting the complexity of the flow models, and thereby balancing goodness of fit against computational cost. Finally, while this paper focuses on binary response data under the M2PL model, the proposed framework can be extended to more complex latent variable models with flexible non-Gaussian latent distributions, including polytomous response models for ordinal or nominal item responses and structured or deep latent variable models.

{\spacingset{1.65}
 \bibstyle{agsm}
\bibliography{references}}

\newpage 
\appendix

{\centering \LARGE \bf Appendix}

\section{Details of the Simulation Design}\label{appendix_11}
In this section, we provide further details on the sparsity patterns in the discrimination parameter. Specifically, let \(A\) be given as 
\[
A=(\alpha_1,\ldots,\alpha_J)^\top \in \mathbb{R}^{J\times K},
\]
where the \(j\)th row corresponds to the discrimination vector
\(\alpha_j^\top\). The nonzero structure of \(A\) is determined by a binary
matrix \(Q=(Q_{jk})\). Conditional on \(Q\), the entries of \(A\) are generated as
\[
A_{jk} =
\begin{cases}
U_{jk}, & Q_{jk}=1,\\
0, & Q_{jk}=0,
\end{cases}
\qquad
U_{jk}\overset{\text{i.i.d.}}{\sim}\mathrm{Unif}(1,2).
\]
Below we explicitly describe the sparse patterns $Q$ for \(K=2,6,10\). 
\begin{itemize}
    \item {\bf \(K=2\).}
Each item loads on one latent dimension. Let $B_2 = I_2$.
The periodic indicator matrix is defined by
\[
Q_{(K=2)} =
\begin{pmatrix}
I_2
B_2\\
B_2\\
\vdots
\end{pmatrix}.
\]
Thus, the rows of \(Q_{(K=2)}\) alternate between \((1,0)\) and \((0,1)\).
\item  {\bf \(K=6\).}
Each non-anchor item loads on two latent dimensions. The first six rows form an identity block. Define
\[
B_6 =
\begin{pmatrix}
1&0&0&1&0&0\\
0&1&0&0&1&0\\
0&0&1&0&0&1
\end{pmatrix}.
\]
Then the periodic sparsity pattern is defined by
\[
Q_{(K=6)}
=
\begin{pmatrix}
I_6\\
B_6\\
B_6\\
\vdots
\end{pmatrix}.
\]
Thus, after the first six anchor items, the sparsity pattern repeats as
\((1,4)\), \((2,5)\), and \((3,6)\). 
\item {\bf \(K=10\).}
Each non-anchor item loads on three latent dimensions. The first ten rows form
an identity block. Define
\[
B_{10} =
\begin{pmatrix}
1&0&0&1&0&0&0&1&0&0\\
0&1&0&0&1&0&0&0&1&0\\
0&0&1&0&0&1&0&0&0&1\\
1&0&0&1&0&0&1&0&0&0\\
0&1&0&0&1&0&0&1&0&0\\
0&0&1&0&0&1&0&0&1&0\\
0&0&0&1&0&0&1&0&0&1\\
1&0&0&0&1&0&0&1&0&0\\
0&1&0&0&0&1&0&0&1&0\\
0&0&1&0&0&0&1&0&0&1
\end{pmatrix}.
\]
Then the periodic sparsity pattern is defined by
\[
Q_{(K=10)}
=
\begin{pmatrix}
I_{10}\\
B_{10}\\
B_{10}\\
\vdots
\end{pmatrix}.
\]
Thus, after the first ten anchor items, the sparsity pattern repeats according to the cyclic offsets \((0,3,7)\). 
\end{itemize}
For all three cases, when \(J=20\) or \(50\), we take the first \(J\) rows of the corresponding periodic matrix as the final indicator matrix \(Q\).

\section{Additional Results on Real Data Analysis}\label{supp_real_data}
\begin{figure}
    \centering
    \captionsetup[subfigure]{font=footnotesize}
    \captionsetup{font=small, skip=3pt}

    \includegraphics[width=0.35\textwidth]{legend_0.png}
    \vspace{-0.5em}

    \begin{subfigure}[b]{0.3\textwidth}
        \centering
        \includegraphics[width=\textwidth]{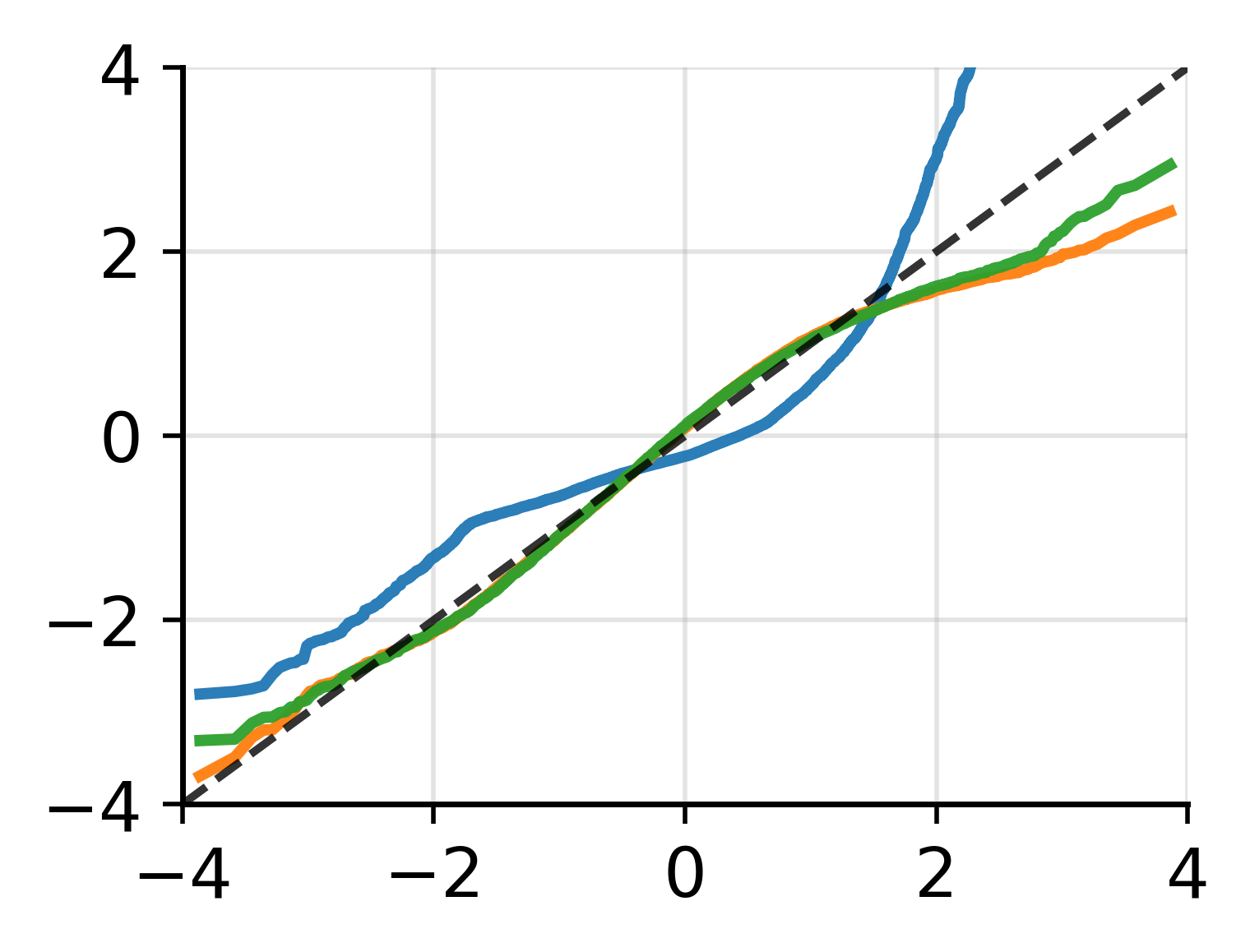}
        \caption{$\theta_1$ (Agreeableness)}
        \label{fig:qq_z1}
    \end{subfigure}
    \hfill
    \begin{subfigure}[b]{0.3\textwidth}
        \centering
        \includegraphics[width=\textwidth]{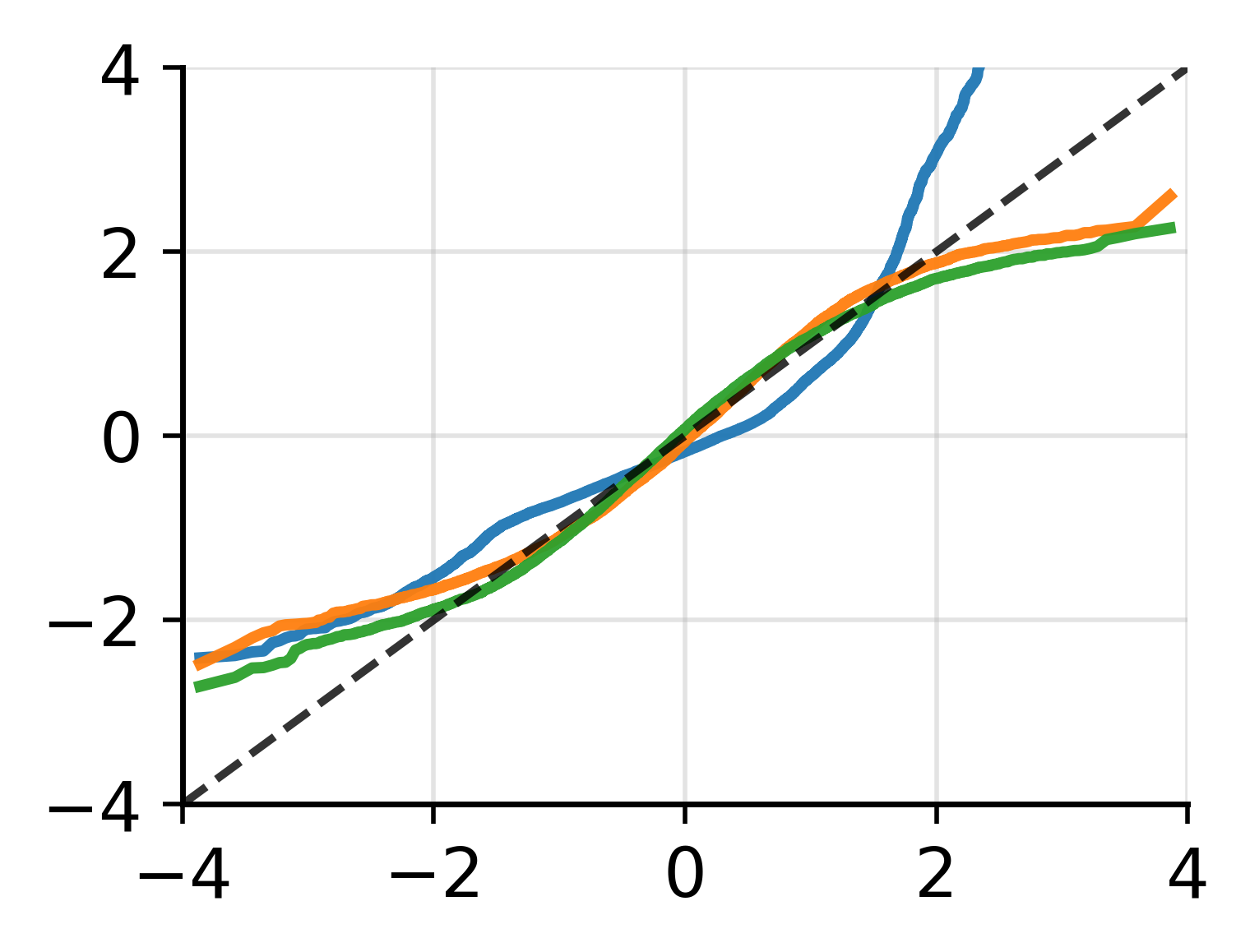}
        \caption{$\theta_2$ (Openness)}
        \label{fig:qq_z2}
    \end{subfigure}
    \hfill
    \begin{subfigure}[b]{0.3\textwidth}
        \centering
        \includegraphics[width=\textwidth]{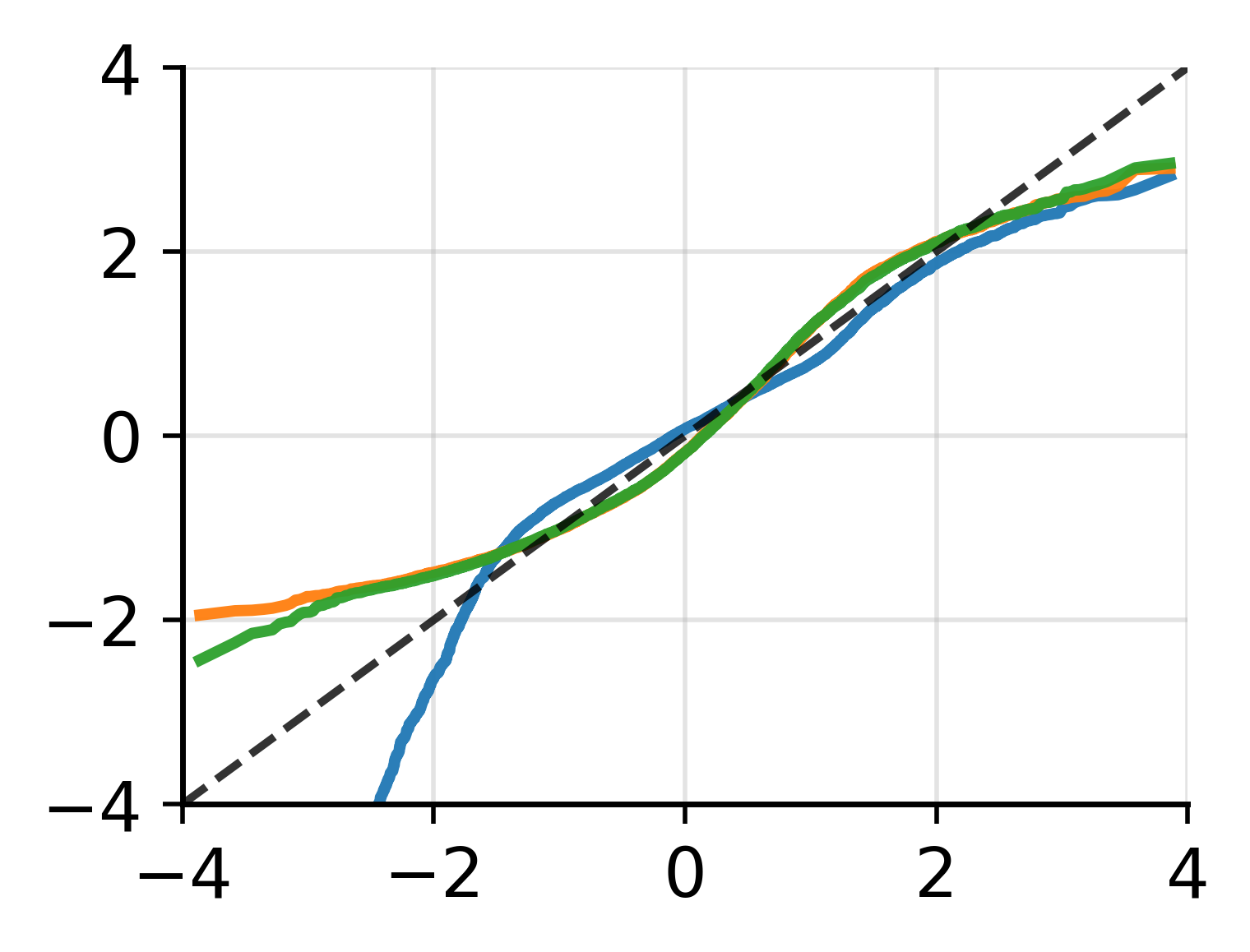}
        \caption{$\theta_3$ (Conscientiousness)}
        \label{fig:qq_z3}
    \end{subfigure}
    \hfill
    \begin{subfigure}[b]{0.3\textwidth}
        \centering
        \includegraphics[width=\textwidth]{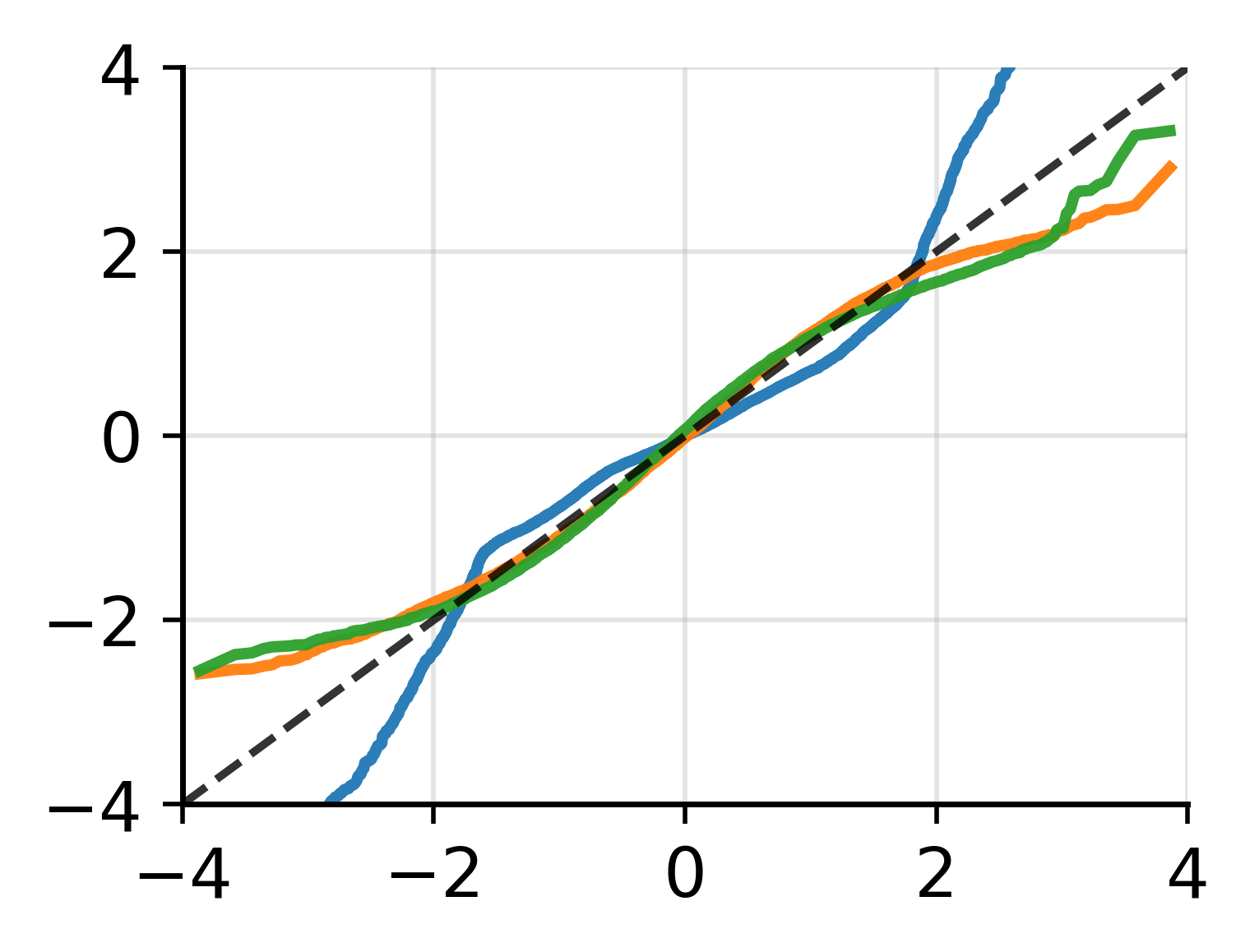}
        \caption{$\theta_4$ (Extraversion)}
        \label{fig:qq_z4}
    \end{subfigure}
    \begin{subfigure}[b]{0.3\textwidth}
        \centering
        \includegraphics[width=\textwidth]{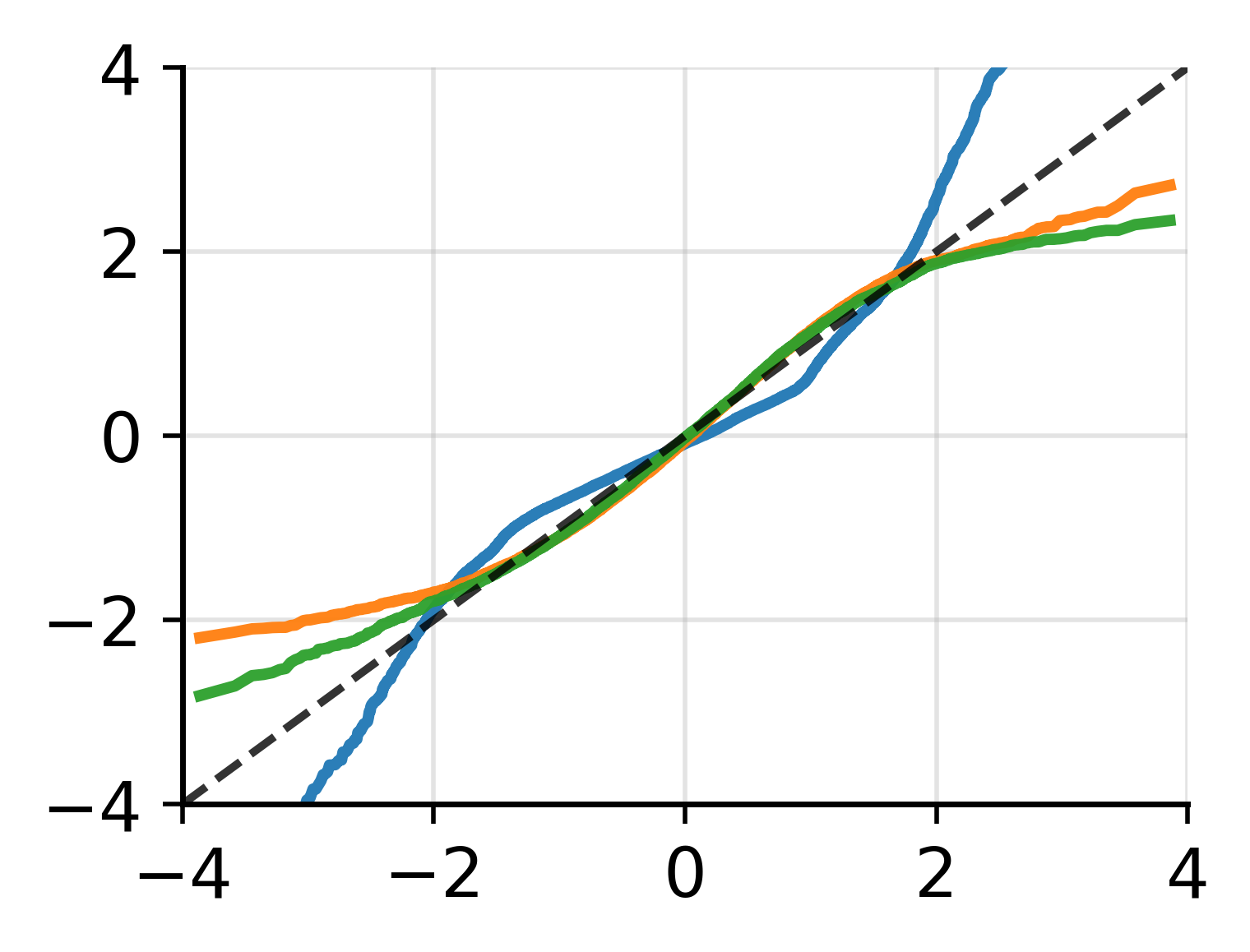}
        \caption{$\theta_5$ (Neuroticism)}
        \label{fig:qq_z5}
    \end{subfigure}

    \vspace{-0.4em}
    \caption{Normal QQ plots of the estimated latent traits under MCEM, MHRM, and our method.}
    \label{fig:qq_latent}
\end{figure}

Figure~\ref{fig:qq_latent} further illustrates deviations from normality in the latent traits estimated by different methods using QQ plots. Across all dimensions, the quantiles estimated by the proposed method show substantial departures from the Gaussian reference line, indicating clear non-Gaussian patterns. In particular, $\theta_1$ and $\theta_2$ exhibit right-skewed behavior, $\theta_3$ exhibits left-skewness, and $\theta_4$ and $\theta_5$ show pronounced deviations in both tails, suggesting heavier-tailed distributions. In contrast, the estimates from MCEM and MHRM remain close to the Gaussian reference line, reflecting the effect of the imposed normality assumption. Since the summed-score likelihood-based test rejects the Gaussian assumption, these two benchmark methods do not adequately capture the complex latent distribution.

\end{document}